\documentclass[12pt]{article}
\usepackage[dvips]{graphicx}
\usepackage{epsfig}
\def\NIMA#1#2#3{Nucl. Inst. Methods {\bf A#1} (#2) #3}

\begin{document}

\begin{center}
EUROPEAN ORGANIZATION FOR NUCLEAR RESEARCH\\
\end{center}
\begin{flushright}
CERN-PH-EP-2009-005\\
2 March 2009
\end{flushright}

\begin{center}
\boldmath {\Large\bf Precise measurement of the $K^\pm\to\pi^\pm
e^+e^-$ decay} \unboldmath
\end{center}

\begin{center}
{\Large The NA48/2 Collaboration}\\
\vspace{2mm}
 J.R.~Batley,
 A.J.~Culling,
 G.~Kalmus,
 C.~Lazzeroni$\,$\footnotemark[1],
 D.J.~Munday,\\
 M.W.~Slater$\,$\footnotemark[1],
 S.A.~Wotton \\
{\em \small Cavendish Laboratory, University of Cambridge,
Cambridge, CB3 0HE, UK$\,$\footnotemark[2]} \\[0.2cm]
 R.~Arcidiacono$\,$\footnotemark[3],
 G.~Bocquet,
 N.~Cabibbo$\,$\footnotemark[4],
 A.~Ceccucci,
 D.~Cundy$\,$\footnotemark[5],
 V.~Falaleev,
 M.~Fidecaro,
 L.~Gatignon,
 A.~Gonidec,
 W.~Kubischta,
 A.~Norton$\,$\footnotemark[6],
 A.~Maier,\\
 M.~Patel,
 A.~Peters\\
{\em \small CERN, CH-1211 Gen\`eve 23, Switzerland} \\[0.2cm]
 S.~Balev$\,$\footnotemark[7],
 P.L.~Frabetti,
 E.~Goudzovski$\,$\renewcommand{\thefootnote}{\fnsymbol{footnote}}%
\footnotemark[1]\renewcommand{\thefootnote}{\arabic{footnote}}\footnotemark[1],
 P.~Hristov$\,$\footnotemark[8],
 V.~Kekelidze,
 V.~Kozhuharov$\,$\footnotemark[9],
 L.~Litov,
 D.~Madigozhin,
 E.~Marinova$\,$\renewcommand{\thefootnote}{\fnsymbol{footnote}}%
\footnotemark[1]\renewcommand{\thefootnote}{\arabic{footnote}}\footnotemark[10],
 N.~Molokanova,
 I.~Polenkevich,\\
 Yu.~Potrebenikov,
 S.~Stoynev$\,$\footnotemark[11],
 A.~Zinchenko \\
{\em \small Joint Institute for Nuclear Research, 141980 Dubna,
Moscow region, Russia} \\[0.2cm]
 E.~Monnier$\,$\footnotemark[12],
 E.~Swallow,
 R.~Winston\\
{\em \small The Enrico Fermi Institute, The University of Chicago,
Chicago, IL 60126, USA}\\[0.2cm]
 P.~Rubin$\,$\footnotemark[13],
 A.~Walker \\
{\em \small Department of Physics and Astronomy, University of
Edinburgh, JCMB King's Buildings, Mayfield Road, Edinburgh, EH9 3JZ, UK} \\[0.2cm]
 W.~Baldini,
 A.~Cotta Ramusino,
 P.~Dalpiaz,
 C.~Damiani,
 M.~Fiorini$\,$\footnotemark[8],
 A.~Gianoli,
 M.~Martini,
 F.~Petrucci,
 M.~Savri\'e,
 M.~Scarpa,
 H.~Wahl \\
{\em \small Dipartimento di Fisica dell'Universit\`a e Sezione
dell'INFN di Ferrara, I-44100 Ferrara, Italy} \\[0.2cm]
 A.~Bizzeti$\,$\footnotemark[14],
 M.~Calvetti,
 E.~Celeghini,
 E.~Iacopini,
 M.~Lenti,
 F.~Martelli$\,$\footnotemark[15],
 G.~Ruggiero$\,$\footnotemark[7],
 M.~Veltri$\,$\footnotemark[15] \\
{\em \small Dipartimento di Fisica dell'Universit\`a e Sezione
dell'INFN di Firenze, I-50125 Firenze, Italy} \\[0.2cm]
 M.~Behler,
 K.~Eppard,
 K.~Kleinknecht,
 P.~Marouelli,
 L.~Masetti$\,$\footnotemark[16],
 U.~Moosbrugger,
 C.~Morales Morales,
 B.~Renk,
 M.~Wache,
 R.~Wanke,
 A.~Winhart \\
{\em \small Institut f\"ur Physik, Universit\"at Mainz, D-55099
 Mainz, Germany$\,$\footnotemark[17]} \\[0.2cm]
 D.~Coward$\,$\footnotemark[18],
 A.~Dabrowski,
 T.~Fonseca Martin$\,$\footnotemark[19],
 M.~Shieh,
 M.~Szleper,\\
 M.~Velasco,
 M.D.~Wood$\,$\footnotemark[20] \\
{\em \small Department of Physics and Astronomy, Northwestern
University, Evanston, IL 60208, USA}\\[0.2cm]
 G.~Anzivino,
 P.~Cenci,
 E.~Imbergamo,
 A.~Nappi,
 M.~Pepe,
 M.C.~Petrucci,\\
 M.~Piccini,
 M.~Raggi$\,$\footnotemark[21],
 M.~Valdata-Nappi \\
{\em \small Dipartimento di Fisica dell'Universit\`a e Sezione
dell'INFN di Perugia, I-06100 Perugia, Italy} \\[0.2cm]
 C.~Cerri,
 R.~Fantechi \\
{\em Sezione dell'INFN di Pisa, I-56100 Pisa, Italy} \\[0.2cm]
 G.~Collazuol,
 L.~DiLella,
 G.~Lamanna,
 I.~Mannelli,
 A.~Michetti \\
{\em Scuola Normale Superiore e Sezione dell'INFN di Pisa, I-56100
Pisa, Italy} \\[0.2cm]
\newpage
 F.~Costantini,
 N.~Doble,
 L.~Fiorini$\,$\footnotemark[22],
 S.~Giudici,
 G.~Pierazzini,\
 M.~Sozzi,
 S.~Venditti \\
{\em Dipartimento di Fisica dell'Universit\`a e Sezione dell'INFN di
Pisa, I-56100 Pisa Italy} \\[0.2cm]
 B.~Bloch-Devaux,
 C.~Cheshkov$\,$\footnotemark[8],
 J.B.~Ch\`eze,
 M.~De Beer,
 J.~Derr\'e,
 G.~Marel,
 E.~Mazzucato,
 B.~Peyaud,
 B.~Vallage \\
{\em \small DSM/IRFU -- CEA Saclay, F-91191 Gif-sur-Yvette, France} \\[0.2cm]
 M.~Holder,
 M.~Ziolkowski \\
{\em \small Fachbereich Physik, Universit\"at Siegen, D-57068
 Siegen, Germany$\,$\footnotemark[23]} \\[0.2cm]
 S.~Bifani$\,$\footnotemark[24],
 C.~Biino,
 N.~Cartiglia,
 M.~Clemencic$\,$\footnotemark[8],
 S.~Goy Lopez$\,$\footnotemark[25],
 F.~Marchetto \\
{\em \small Dipartimento di Fisica Sperimentale dell'Universit\`a e
Sezione dell'INFN di Torino,\\ I-10125 Torino, Italy} \\[0.2cm]
 H.~Dibon,
 M.~Jeitler,
 M.~Markytan,
 I.~Mikulec,
 G.~Neuhofer,
 L.~Widhalm \\
{\em \small \"Osterreichische Akademie der Wissenschaften, Institut
f\"ur Hochenergiephysik,\\ A-10560 Wien, Austria$\,$\footnotemark[26]} \\[0.5cm]
\it{Accepted for publication in Physics Letters B.} \rm
\end{center}

\setcounter{footnote}{0}
\renewcommand{\thefootnote}{\fnsymbol{footnote}}
\footnotetext[1]{Corresponding authors, emails:
eg@hep.ph.bham.ac.uk, evelina.marinova@cern.ch}
\renewcommand{\thefootnote}{\arabic{footnote}}
\footnotetext[1]{University of Birmingham, Edgbaston, Birmingham,
B15 2TT, UK}
\footnotetext[2]{Funded by the UK Particle Physics and Astronomy
Research Council}
\footnotetext[3]{Dipartimento di Fisica Sperimentale
dell'Universit\`a e Sezione dell'INFN di Torino, I-10125 Torino,
Italy}
\footnotetext[4]{Universit\`a di Roma ``La Sapienza'' e Sezione
dell'INFN di Roma, I-00185 Roma, Italy}
\footnotetext[5]{Istituto di Cosmogeofisica del CNR di Torino,
I-10133 Torino, Italy}
\footnotetext[6]{Dipartimento di Fisica dell'Universit\`a e Sezione
dell'INFN di Ferrara, I-44100 Ferrara, Italy}
\footnotetext[7]{Scuola Normale Superiore, I-56100 Pisa, Italy}
\footnotetext[8]{CERN, CH-1211 Gen\`eve 23, Switzerland}
\footnotetext[9]{Faculty of Physics, University of Sofia ``St. Kl.
Ohridski'', 5 J. Bourchier Blvd., 1164 Sofia, Bulgaria}
\footnotetext[10]{Sezione dell'INFN di Perugia, I-06100 Perugia,
Italy}
\footnotetext[11]{Northwestern University, 2145 Sheridan Road,
Evanston, IL 60208, USA}
\footnotetext[12]{Centre de Physique des Particules de Marseille,
IN2P3-CNRS, Universit\'e de la M\'editerran\'ee, Marseille, France}
\footnotetext[13]{Department of Physics and Astronomy, George Mason
University, Fairfax, VA 22030, USA}
\footnotetext[14]{Istituto di Fisica, Universit\`a di Modena e
Reggio Emilia, I-41100 Modena, Italy}
\footnotetext[15]{Istituto di Fisica, Universit\`a di Urbino,
I-61029 Urbino, Italy}
\footnotetext[16]{Physikalisches Institut, Universit\"at Bonn,
D-53115 Bonn, Germany}
\footnotetext[17]{Funded by the German Federal Minister for
Education and research under contract 05HK1UM1/1}
\footnotetext[18]{SLAC, Stanford University, Menlo Park, CA 94025,
USA}
\footnotetext[19]{Royal Holloway, University of London, Egham Hill,
Egham, TW20 0EX, UK}
\footnotetext[20]{UCLA, Los Angeles, CA 90024, USA}
\footnotetext[21]{Laboratori Nazionali di Frascati, via E. Fermi,
40, I-00044 Frascati (Rome), Italy}
\footnotetext[22]{Institut de F\'isica d'Altes Energies, UAB,
E-08193 Bellaterra (Barcelona), Spain}
\footnotetext[23]{Funded by the German Federal Minister for Research
and Technology (BMBF) under contract 056SI74}
\footnotetext[24]{University of Bern, Institute for Theoretical
Physics, Sidlerstrasse 5, CH-3012 Bern, Switzerland}
\footnotetext[25]{Centro de Investigaciones Energeticas
Medioambientales y Tecnologicas, E-28040 Madrid, Spain}
\footnotetext[26]{Funded by the Austrian Ministry for Traffic and
Research under the contract GZ 616.360/2-IV GZ 616.363/2-VIII, and
by the Fonds f\"ur Wissenschaft und Forschung FWF Nr.~P08929-PHY}

%%%%%%%%%%%%%%%%%%%%%%%%%%%%%%%%%%%%%%%%%%%%%%%%%%%%%%
%%%\begin{linenumbers}

\newpage
\begin{abstract}
A sample of 7253 $K^\pm\to\pi^\pm e^+e^-(\gamma)$ decay candidates
with 1.0\% background contamination has been collected by the NA48/2
experiment at the CERN SPS, which allowed a precise measurement of
the decay properties. The branching ratio in the full kinematic
range was measured to be ${\rm BR}=(3.11\pm0.12)\times 10^{-7}$,
where the uncertainty includes also the model dependence. The shape
of the form factor $W(z)$, where $z=(M_{ee}/M_K)^2$, was
parameterized according to several models, and, in particular, the
slope $\delta$ of the linear form factor $W(z)=W_0(1+\delta z)$ was
determined to be $\delta=2.32\pm0.18$. A possible CP violating
asymmetry of $K^+$ and $K^-$ decay widths was investigated, and a
conservative upper limit of $2.1\times 10^{-2}$ at 90\% CL was
established.
\end{abstract}

\newpage

%%%%%%%%%%%%%%%%%%%%%%%%%%%%%%%%%%%%%%%
\section*{Introduction}

Radiative nonleptonic kaon decays represent a source of information
on the structure of the weak interactions at low energies. The
flavour-changing neutral current process $K^\pm\to\pi^\pm e^+e^-$,
induced at one-loop level in the Standard Model and highly
suppressed by the GIM mechanism~\cite{gim}, is of particular
interest. The $K^\pm\to\pi^\pm l^+l^-$ processes have been described
by the Chiral Perturbation Theory (ChPT)~\cite{ek87}; several models
predicting the form factor characterizing the dilepton invariant
mass spectrum, and thus the decay rate, have been
proposed~\cite{da98,fr04,du06}.

The first observation of the $K^+\to\pi^+e^+e^-$ process was made at
the CERN PS more than 30 years ago~\cite{bl75}, followed more
recently by BNL E777~\cite{al92} and E865~\cite{ap99} measurements.
The most precise of these, E865, based on a sample of 10300
candidates with 1.2\% background, allowed a detailed analysis of the
decay form factor and rate, and a test of the next-to-leading order
ChPT calculation~\cite{da98}.

In this paper, a new precise measurement of the $K^\pm\to\pi^\pm
e^+e^-$ decay is reported, based on the full data set collected in
2003--2004 by the NA48/2 experiment at the CERN SPS. In addition to
spectrum and rate studies, the first observation of the $K^-$ decay
allowed setting an upper limit on the charge asymmetry of
$K^+\to\pi^+e^+e^-$ and $K^-\to\pi^-e^+e^-$ decay rates, which can
be related to CP violation.

%%%%%%%%%%%%%%%%%%%%%%%%%%%%%%%%%%%%%%%%%%%%%%%%%
\section{The NA48/2 experiment}

\begin{figure}[tb]
\vspace{-6mm}
\begin{center}
{\resizebox*{\textwidth}{!}{\includegraphics{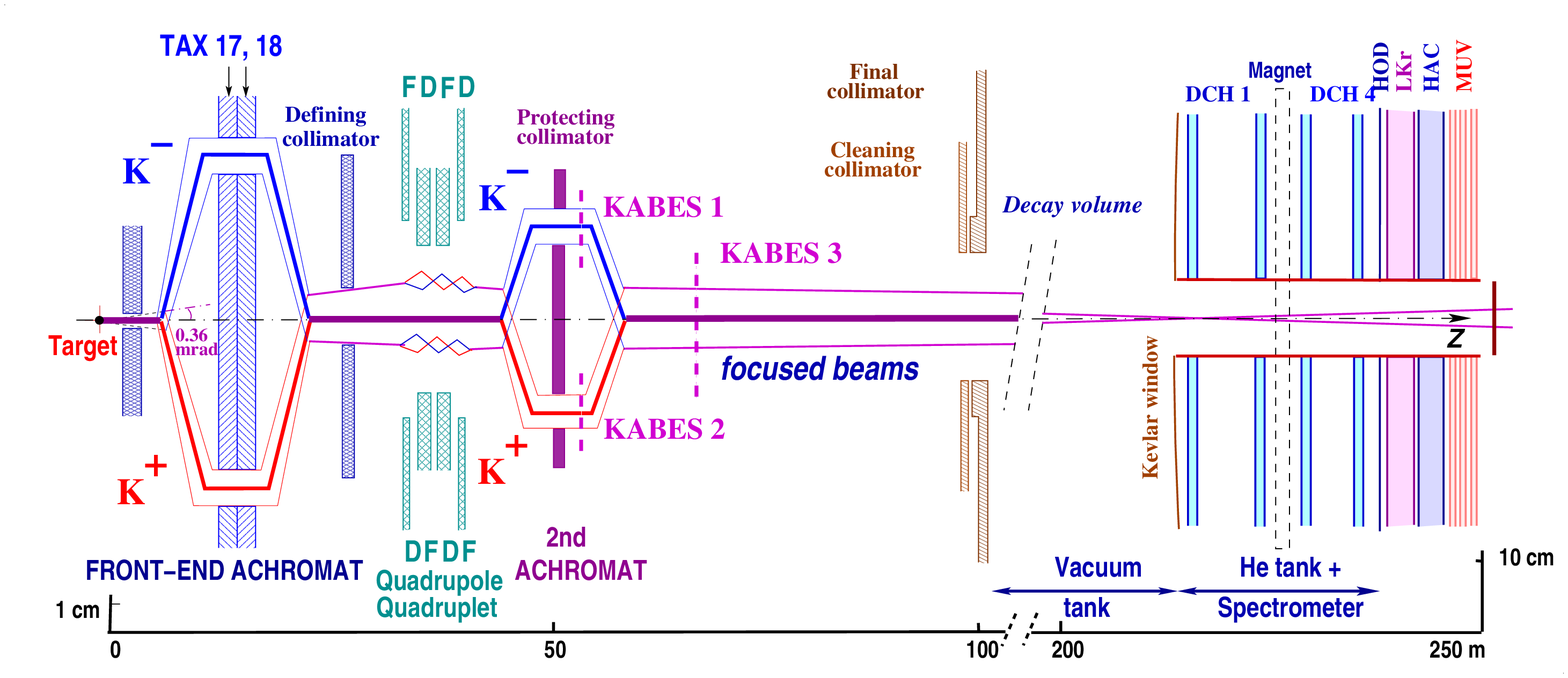}}}
\end{center}
\vspace{-6mm} \caption{Schematic lateral view of the NA48/2 beam
line (TAX17,18: motorized beam dump/collimators used to select the
momentum of the $K^+$ and $K^-$ beams; FDFD/DFDF: focusing set of
quadrupoles, KABES1--3: beam spectrometer stations not used in this
analysis), decay volume, and detector (DCH1--4: drift chambers, HOD:
hodoscope, LKr: EM calorimeter, HAC: hadron calorimeter, MUV: muon
veto). Note that the vertical scales are different in the two parts
of the figure.} \label{fig:beams}
\end{figure}

The NA48/2 experiment, specifically designed for charge asymmetry
measurements~\cite{ba07}, uses simultaneous $K^+$ and $K^-$ beams
produced by 400 GeV/$c$ primary SPS protons impinging on a beryllium
target. Charged particles with momentum $(60\pm3)$ GeV/$c$ are
selected by an achromatic system of four dipole magnets with zero
total deflection (`achromat'), which splits the two beams in the
vertical plane and then recombines them on a common axis. The beams
pass through momentum defining collimators and a series of four
quadrupoles designed to focus the beams at the detector. Finally the
two beams are again split in the vertical plane and recombined in a
second achromat. The layout of the beams and detectors is shown
schematically in Fig.~\ref{fig:beams}.

The beams then enter the fiducial decay volume housed in a 114 m
long cylindrical vacuum tank with a diameter of 1.92 m, extended to
2.4 m after the first 65 m. Both beams follow the same path in the
decay volume: their axes coincide within 1~mm, while the transverse
size of the beams is about 1~cm. With $7\times 10^{11}$ protons
incident on the target per SPS spill of $4.8$~s duration, the
positive (negative) beam flux at the entrance of the decay volume is
$3.8\times 10^7$ ($2.6\times 10^7$) particles per pulse, of which
$5.7\%$ ($4.9\%$) are $K^+$ ($K^-$). The $K^+/K^-$ flux ratio is
1.79. The fraction of beam kaons decaying in the decay volume at
nominal momentum is $22\%$.

A detailed description of the NA48 detector can be found
in~\cite{fa07}. The decay volume is followed by a magnetic
spectrometer housed in a tank filled with helium at nearly
atmospheric pressure, separated from the vacuum tank by a thin
($0.31\%X_0$) Kevlar composite window. A thin-walled aluminium beam
pipe of 16~cm outer diameter traversing the centre of the
spectrometer (and all the following detectors) allows the undecayed
beam particles and the muon halo from decays of beam pions to
continue their path in vacuum. The spectrometer consists of four
drift chambers (DCH): DCH1, DCH2 located upstream, and DCH3, DCH4
downstream of a dipole magnet. The magnet provides a horizontal
transverse momentum kick $\Delta p=120~{\rm MeV}/c$ for charged
particles. Each DCH is composed of eight planes of sense wires. The
spatial resolution of each DCH is $\sigma_x=\sigma_y=90~\mu$m. The
nominal spectrometer momentum resolution is $\sigma_p/p = (1.02
\oplus 0.044\cdot p)\%$ ($p$ in GeV/$c$).

The magnetic spectrometer is followed by a plastic scintillator
hodoscope (HOD) used to produce fast trigger signals and to provide
precise time measurements of charged particles. The hodoscope has a
regular octagonal shape with a transverse size of about 2.4~m, and
consists of a plane of horizontal and a plane of vertical
strip-shaped counters.

The HOD is followed by a liquid krypton electromagnetic calorimeter
(LKr) used for photon detection and particle identification. It is
an almost homogeneous ionization chamber with an active volume of 7
m$^3$ of liquid krypton, segmented transversally into 13248
projective cells, 2$\times$2 cm$^2$ each, and with no longitudinal
segmentation. The calorimeter is $27X_0$ deep and has an energy
resolution $\sigma(E)/E=0.032/\sqrt{E}\oplus0.09/E\oplus0.0042$ ($E$
in GeV). Spatial resolution for an isolated electromagnetic shower
is $\sigma_x=\sigma_y=0.42/\sqrt{E}\oplus0.06$ cm ($E$ in GeV) for
the transverse coordinates $x$ and $y$.

The LKr is followed by a hadronic calorimeter (HAC) and a muon
detector (MUV), both not used in the present analysis.

%%%%%%%%%%%%%%%%%%%%%%%%%%%%%%%%
\section{Data analysis}

\noindent The $K^\pm\to\pi^\pm e^+e^-$ rate is measured relative to
the more abundant $K^\pm\to\pi^\pm\pi^0_D$ normalisation channel
(where $\pi^0_D\to e^+e^-\gamma$ is the so called Dalitz decay). The
final states of the signal and normalisation channels contain
identical sets of charged particles. Thus electron and pion
identification efficiencies, potentially representing a significant
source of systematic uncertainties, cancel in the first order.

\vspace{2mm}

\noindent {\bf Monte Carlo simulation} \vspace{2mm}

\noindent In order to compute acceptances for signal, normalisation
and background channels, a detailed GEANT3-based~\cite{geant} Monte
Carlo (MC) simulation is employed, which includes full detector
geometry and material description, stray magnetic fields, DCH local
inefficiencies and misalignment, detailed simulation of the kaon
beam line, and time variations of the above throughout the running
period.

\vspace{2mm}

\noindent {\bf Event selection}

\vspace{2mm}

\noindent Three-track vertices (compatible with the topology of
$K^\pm\to\pi^\pm e^+e^-$ and $K^\pm\to\pi^\pm\pi^0_D$ decays) are
reconstructed by extrapolation of track segments from the upstream
part of the spectrometer back into the decay volume, taking into
account the measured Earth's magnetic field, stray fields due to
magnetization of the vacuum tank, and multiple scattering.

\begin{figure}[tb]
\vspace{-2mm}
\begin{center}
{\resizebox*{0.5\textwidth}{!}{\includegraphics{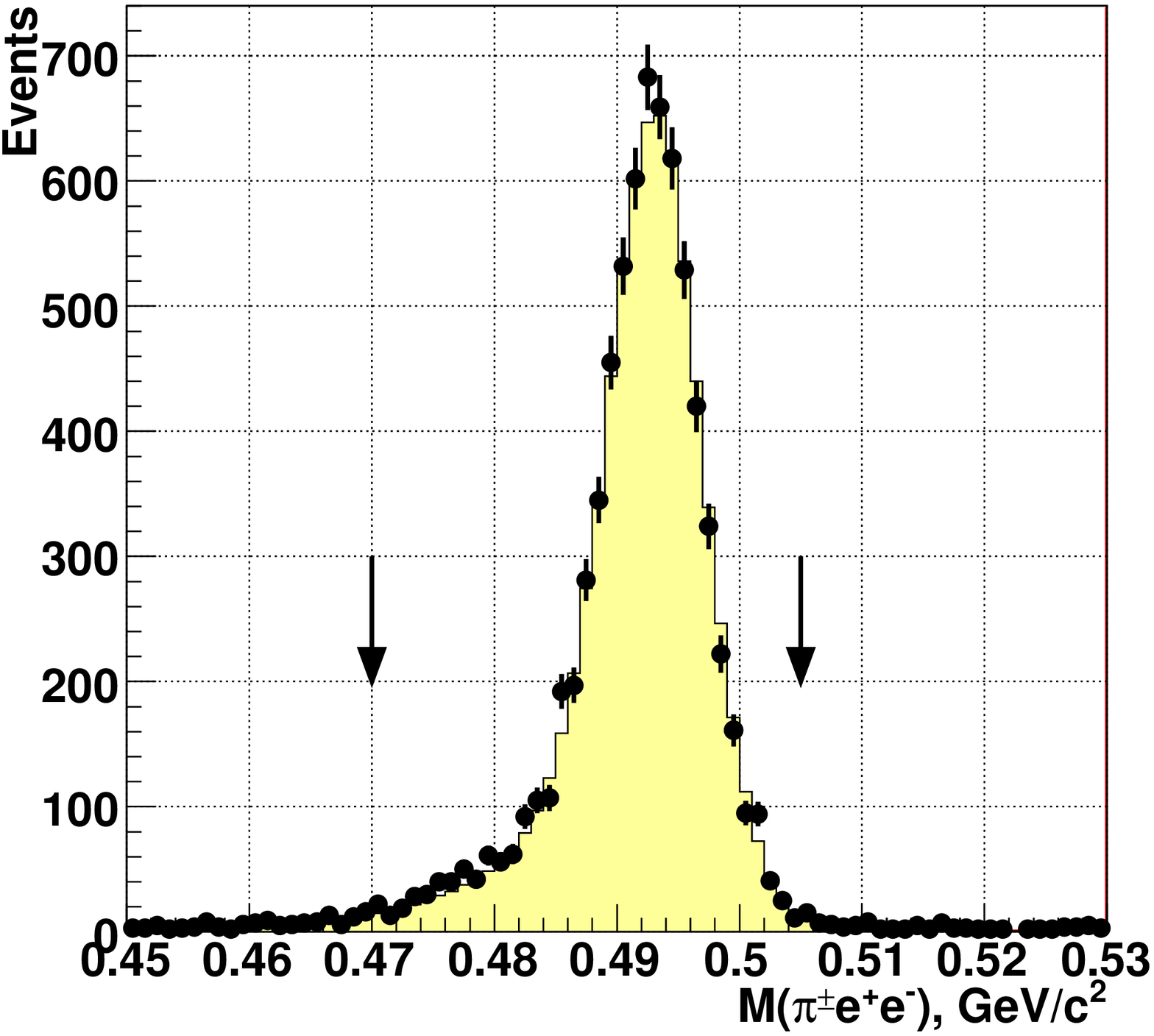}}}%
{\resizebox*{0.5\textwidth}{!}{\includegraphics{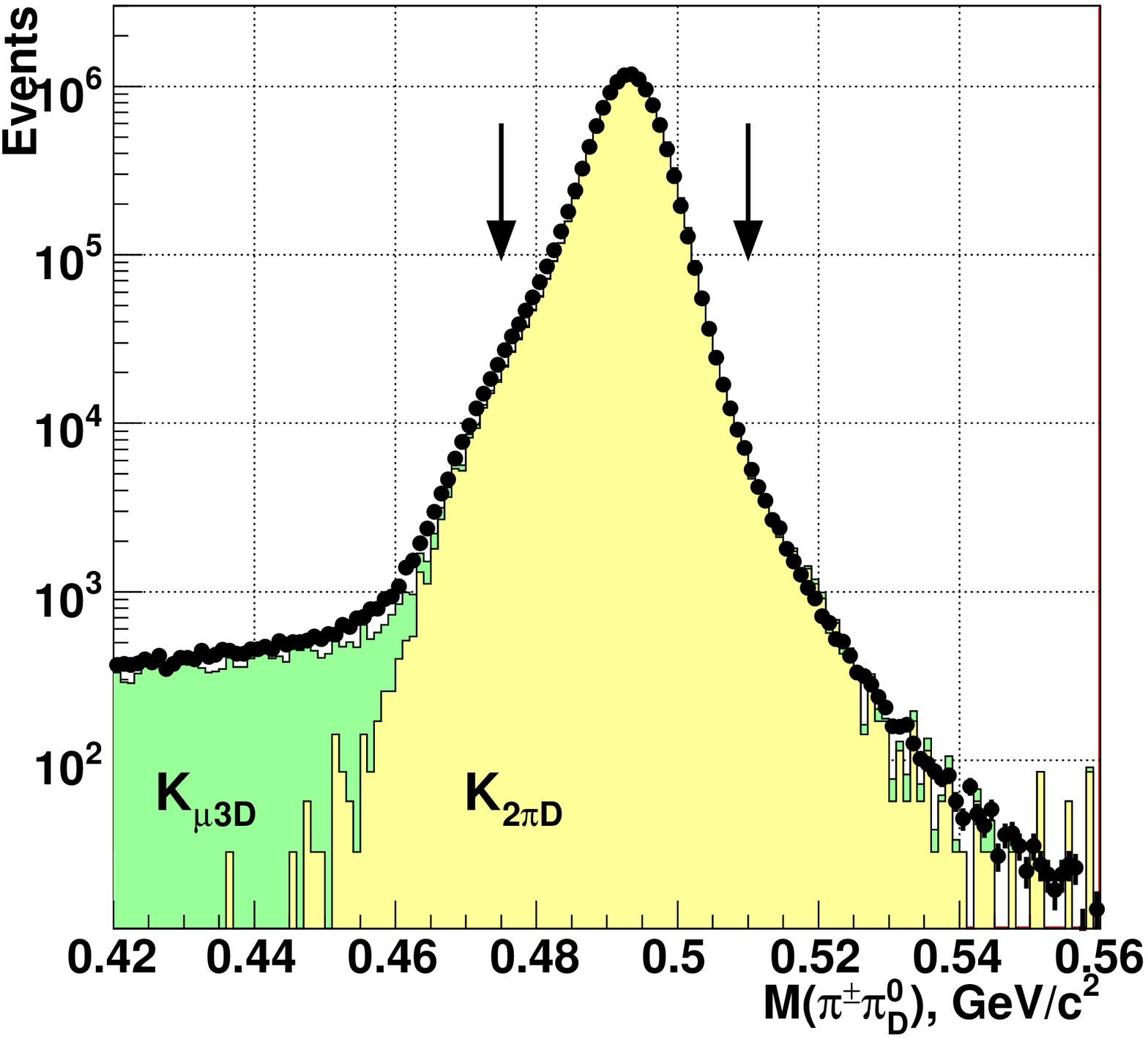}}}
\put(-261,177){\bf\large (a)} \put(-35,177){\bf\large (b)}
\end{center}
\vspace{-8mm} \caption{(a) Reconstructed spectrum of $\pi^\pm
e^+e^-$ invariant mass: data (dots) and MC simulation (filled area).
Note the description of the radiative mass tail by the PHOTOS
simulation. (b) Reconstructed spectrum of $\pi^\pm\pi^0_D$ invariant
mass: data (dots) and MC simulations of $K^\pm\to\pi^\pm\pi^0_D$
signal and $K^\pm\to\pi^0_D\mu^\pm\nu$ background (filled areas).
Signal regions are indicated with arrows.} \label{fig:mk}
\end{figure}

A large part of the selection is common to the signal and
normalisation modes. It requires the presence of a vertex satisfying
the following criteria.
\begin{itemize}
\item Vertex longitudinal position is inside fiducial decay volume:
$Z_{\rm vertex}>Z_{\rm final~collimator}$.
\item The vertex tracks are required to be consistent in time
(within a 10~ns time window) and consistent with the trigger time,
to be in DCH, HOD and LKr geometric acceptance, and to have momenta
in the range $5~{\rm GeV}/c<p<50~{\rm GeV}/c$. Track separations are
required to exceed 2~cm in the DCH1 plane to suppress photon
conversions, and to exceed 15~cm in the LKr front plane to minimize
particle misidentification due to shower overlaps.
\item Total charge of the three tracks: $Q=\pm1$.
\item Particle identification is performed using the ratio $E/p$ of
energy deposition of the particle in the LKr calorimeter to its
momentum measured by the spectrometer. The vertex is required to be
composed of one $\pi$ candidate ($E/p<0.85$), and a pair of
oppositely charged $e^\pm$ candidates ($E/p>0.95$). No
discrimination of pions against muons is performed.
\end{itemize}
If several vertices satisfy the above conditions, the one with the
best vertex fit quality is considered. The $K^\pm\to\pi^\pm e^+e^-$
candidates are selected by applying the following criteria to the
reconstructed kinematic variables.
\begin{itemize}
\item $\pi^\pm e^+e^-$ momentum within the beam nominal range:
$54~{\rm GeV}/c<|\vec p_{\pi ee}|<66~{\rm GeV}/c$.
\item $\pi^\pm e^+e^-$ transverse momentum with respect to
the beam trajectory (which is precisely measured using the the
concurrently acquired $K^\pm\to3\pi^\pm$ sample): $p_T^2<0.5\times
10^{-3}~({\rm GeV}/c)^2$.
\item Kinematic suppression of the main background channel
$K^\pm\to\pi^\pm\pi^0_D$ (and other minor backgrounds induced by
$\pi^0_D$ and $\pi^0_{DD}\to4e^\pm$ decays) by requiring the
$e^+e^-$ mass to be above the $\pi^0$ mass: $z=(M_{ee}/M_K)^2>0.08$,
which approximately corresponds to $M_{ee}>140$~MeV/$c^2$, and leads
to an unavoidable loss of $\sim 30\%$ of the signal sample.
\item $\pi^\pm e^+e^-$ invariant mass:
$470~{\rm MeV}/c^2<M_{\pi ee}<505~{\rm MeV}/c^2$. The signal region,
asymmetric with respect to the nominal kaon mass
(Fig.~\ref{fig:mk}a), includes a part of the radiative mass tail.
The lower mass limit of $470~{\rm MeV}/c^2$ corresponds to an upper
cutoff $E_\gamma<23.1~{\rm MeV}$ for the energy of a single directly
undetectable soft radiative photon, and is chosen as a trade-off
between making an inclusive analysis, and keeping the background
reasonably small.
\end{itemize}
No restrictions are applied to the $K^\pm\to\pi^\pm e^+e^-$ sample
on the additional energy deposition in the LKr calorimeter, the
reason being two-fold: 1) to avoid bias due to the radiative
$K^\pm\to\pi^\pm e^+e^-\gamma$ decays (with predominantly soft
photons, which makes the LKr response difficult to simulate); 2) to
decrease sensitivity to accidental activity.

For the $K^\pm\to\pi^\pm\pi^0_D$ normalisation mode candidates, the
presence of a LKr energy deposition cluster (photon candidate)
satisfying the following principal criteria is required.
\begin{itemize}
\item Reconstructed cluster energy $E>3$~GeV,
cluster time consistent with the vertex time, sufficient transverse
separations from track impact points at the LKr plane
($R_{\pi\gamma}>30$~cm, $R_{e\gamma}>10$~cm).
\item $e^+e^-\gamma$ invariant mass compatible with a $\pi^0_D$
decay: $|M_{ee\gamma}-M_{\pi^0}|<10$~MeV/$c^2$.
\item The same conditions on reconstructed $\pi^\pm e^+e^-\gamma$
total and transverse momenta as used for $\pi^\pm e^+e^-$ momentum
in the $K^\pm\to\pi^\pm e^+e^-$ selection.
\item $\pi^\pm e^+e^-\gamma$ invariant mass:
$475~{\rm MeV}/c^2<M_{\pi ee\gamma}<510~{\rm MeV}/c^2$.
\end{itemize}

\newpage
\noindent {\bf Signal sample}

\vspace{2mm}

\noindent The reconstructed $\pi^\pm e^+e^-$ invariant mass spectrum
is presented in Fig.~\ref{fig:mk}a. The measured $M_{\pi ee}$
resolution is $\sigma_{\pi ee}=4.2$~MeV/$c^2$, in agreement with MC
simulation. The $e^+e^-$ invariant mass resolution computed by MC
simulation is $\sigma_{ee}=2.3$~MeV/$c^2$. The number of
$K^\pm\to\pi^\pm e^+e^-(\gamma)$ decay candidates in the signal
region is $N_{\pi ee}=7253$, out of which 4613 (2640) are $K^+$
($K^-$) candidates. The residual background sources after the
kinematical suppression of the $\pi^0_D$ and $\pi^0_{DD}$ decays are
the following.
\begin{itemize}
\item Kaon decays with a $\pi^0$ in the final state
followed by a $\pi^0_D\to e^+e^-\gamma$ decay and particle
misidentification ($e^\pm$ identified as $\pi^\pm$ or vice versa).
Two dominant contributions identified with MC simulations are (1)
$K^\pm\to\pi^\pm\pi^0_D$ with misidentified $e^\pm$ and $\pi^\pm$;
(2) $K^\pm\to\pi^0_De^\pm\nu$ with a misidentified $e^\pm$ from the
$\pi^0_D$ decay. The two contributions are of similar size, which
means that about 75\% of electron candidates in this sample are
genuine electrons, and the rest are misidentified pions.
\item Kaon decays with two $e^+e^-$ pairs in the final state
resulting from either $\pi^0_{D(D)}$ decays or external $\gamma$
conversions, only a single lepton from each $e^+e^-$ pair
constituting the best vertex, without particle misidentification
involved. All electron candidates in this sample are genuine
electrons.
\end{itemize}
Backgrounds can be reliably estimated and subtracted using the data
sample itself. For the first of the above background types, the
expected kinematic distribution of the lepton number violating
``same lepton sign'' $\pi^\mp e^\pm e^\pm$ candidates is identical
to that of background events (up to a negligible acceptance
correction). For the second background type, the expected sum of
distributions of the same lepton sign candidates and the $\pi^\pm
e^\pm e^\pm$ candidates with the unphysical total charge $|Q|=3$ is
similarly identical to that of background events. Background
contamination is estimated using the sum of the numbers of same
lepton sign candidates $N_{ss}=55$, and $|Q|=3$ candidates
$N_{q3}=16$ as $(N_{ss}+N_{q3})/N_{\pi ee}=(1.0\pm0.1_{\rm
stat.})\%$.

A cross check of the composition of same lepton sign and $|Q|=3$
samples using a tighter electron identification (based not only on
$E/p$ but also on shower properties and track-cluster matching)
confirms the expected fractions of genuine electrons among the
electron candidates.

\vspace{2mm}

\noindent {\bf Normalisation sample}

\vspace{2mm}

\noindent The reconstructed $\pi^\pm e^+e^-\gamma$ invariant mass
spectrum is presented in Fig.~\ref{fig:mk}b. The number of
$K^\pm\to\pi^\pm\pi^0_D$ candidates in the signal region is
$N_{2\pi}=1.212\times 10^7$, which corresponds to the number of kaon
decays in the fiducial decay volume of about $N_K=1.7\times
10^{11}$. The only significant background source is the semileptonic
$K^\pm\to\pi^0_D\mu^\pm\nu$ decay. Its contribution is not
suppressed by particle identification requirements, since no
$\pi$/$\mu$ separation is performed. The background contamination in
the signal region is estimated to be 0.15\% by MC simulation.

\vspace{2mm}

\noindent {\bf Trigger chain and its efficiency} \vspace{2mm}

\noindent Both $K^\pm\to\pi^\pm e^+e^-$ and $K^\pm\to\pi^\pm\pi^0_D$
samples are recorded via a two-level trigger chain designed to
collect the $K^\pm\to3\pi^\pm$ decays. At the first level (L1), the
HOD surface is logically subdivided into 16 non-overlapping square
regions; a coincidence of hits in the two planes of the HOD is
required to occur in at least two such regions. The second level
(L2) is based on a hardware system computing coordinates of hits
from DCH drift times, and a farm of asynchronous processors
performing fast track reconstruction and running a selection
algorithm, which basically requires the presence of at least two
tracks consistent with a common origin in the decay volume (closest
distance of approach less than 5 cm). L1 triggers not satisfying
this condition are examined further and accepted nevertheless if
there is a reconstructed track not kinematically compatible with a
$\pi^\pm\pi^0$ decay of a $K^\pm$ having momentum of 60 GeV/$c$
directed along the beam axis.

The NA48/2 analysis strategy for non-rare decay modes involves
direct measurement of the trigger efficiencies using control data
samples of downscaled low bias triggers collected simultaneously
with the main triggers. However direct measurements are not possible
for the $K^\pm\to\pi^\pm e^+e^-$ events due to very limited sizes of
the corresponding control samples. Dedicated simulations of L1 and
L2 performance were used instead. The sources of the trigger
inefficiency are the local inefficiencies of the HOD and the DCHs,
which were mapped using special muon runs and the
$K^\pm\to\pi^\pm\pi^+\pi^-$ data, and included into the simulations.
The simulated efficiencies and their kinematic dependencies were
compared against measurements for the abundant
$K^\pm\to\pi^\pm\pi^0_D$ and $K^\pm\to\pi^\pm\pi^+\pi^-$ decays in
order to validate the simulations; simulated values of trigger
efficiencies agree with the measurements to a level of a few
$10^{-4}$ units.

The simulated values of L1 and L2 inefficiencies for the selected
$K^\pm\to\pi^\pm\pi^0_D$ sample are $\varepsilon_{L1}=0.37\%$,
$\varepsilon_{L2}=0.80\%$. The values of the integral trigger
inefficiencies for the $K^\pm\to\pi^\pm e^+e^-$ sample depend on the
a priori unknown form factor; the corrections are applied
differentially in bins of dilepton invariant mass $M_{ee}$.
Indicative values of inefficiencies for a realistic linear form
factor with a slope $\delta=2.3$ are $\varepsilon_{L1}=0.06\%$,
$\varepsilon_{L2}=0.42\%$. The $K^\pm\to\pi^\pm\pi^0_D$ sample is
affected by a larger trigger inefficiency due to the smaller
invariant masses $M_{ee}$, which means that the lepton trajectories
are geometrically closer. Similarly, the inefficiency is enhanced
for $K^\pm\to\pi^\pm e^+e^-$ events at low $M_{ee}$.

\vspace{2mm}

\noindent {\bf Theoretical input} \vspace{2mm}

\noindent The decay is supposed to proceed through single virtual
photon exchange, resulting in a spectrum of the $z=(M_{ee}/M_K)^2$
kinematic variable sensitive to the form factor $W(z)$~\cite{ek87}:
\begin{equation}
\frac{d\Gamma}{dz}=\frac{\alpha^2M_K}{12\pi(4\pi)^4}
\lambda^{3/2}(1,z,r_\pi^2)\sqrt{1-4\frac{r_e^2}{z}}
\left(1+2\frac{r_e^2}{z}\right)|W(z)|^2, \label{theory}
\end{equation}
where $r_e=m_e/M_K$, $r_\pi=m_\pi/M_K$, and
$\lambda(a,b,c)=a^2+b^2+c^2-2ab-2ac-2bc$. On the other hand, the
spectrum of the angle $\theta_{\pi e}$ between $\pi$ and $e^+$ in
the $e^+e^-$ rest frame is $d\Gamma/d\theta_{\pi
e}=C\sin^2\theta_{\pi e}$, $C={\rm const}$, and is not sensitive to
$W(z)$. The expression for two-dimensional decay density can be
found for instance in~\cite{du06}.

The following parameterizations of the form factor $W(z)$ are
considered in the present analysis.
\begin{enumerate}
\item Linear: $W(z)=G_FM_K^2f_0(1+\delta z)$
with free normalisation and slope $(f_0,\delta)$. Decay rate and $z$
spectrum are sensitive to $|f_0|$, not to its sign.
\item Next-to-leading order ChPT~\cite{da98}:
$W(z)=G_FM_K^2(a_++b_+z)+W^{\pi\pi}(z)$ with free parameters
$(a_+,b_+)$ and an explicitly calculated pion loop term
$W^{\pi\pi}(z)$ given in~\cite{da98}.
\item Combined framework of ChPT and large-$N_c$ QCD~\cite{fr04}:
the form factor is parameterized as $W(z)\equiv W(\tilde{\rm
w},\beta,z)$ with free parameters $(\tilde{\rm w},\beta)$, see also
the Appendix.
\item ChPT parameterization~\cite{du06} involving meson form
factors: $W(z)\equiv W(M_a,M_\rho,z)$. The resonance masses ($M_a$,
$M_\rho$) are treated as free parameters in the present analysis.
\end{enumerate}
The goal of the analysis is measuring the form factor parameters in
the framework of each of the above models, and the computation of
the corresponding branching ratios ${\rm BR}_{1,2,3,4}$ by
integration of (\ref{theory}) normalised to full $K^\pm$ decay width
$\Gamma_K$~\cite{pdg}.

The Coulomb correction factor is taken into account
following~\cite{is08} (eq. 20). The effects of radiative corrections
to the $K^\pm\to\pi^\pm e^+e^-$ process are evaluated using the
PHOTOS~\cite{photos} simulation of the
$K^\pm\to\pi^\pm\gamma^*\to\pi^\pm e^+e^-$ decay chain. The results
are cross-checked using a generalized computation for a multi-body
meson decay~\cite{is08} (eq. 19). In addition to affecting the
values of the model parameters, radiative corrections are crucial
for the extrapolation of the branching ratio from the limited
$M_{\pi ee}$ (equivalently, $E_\gamma$) signal region to the full
kinematic region; according to the above models for radiative
corrections, about 6\% of the total $K^\pm\to\pi^\pm e^+e^-(\gamma)$
decay rate fall outside the signal region $E_\gamma<23.1$~MeV.

\vspace{2mm}

\noindent {\bf Fitting procedure} \vspace{2mm}

\begin{figure}[t]
\vspace{-2mm}
\begin{center}
{\resizebox*{0.5\textwidth}{!}{\includegraphics{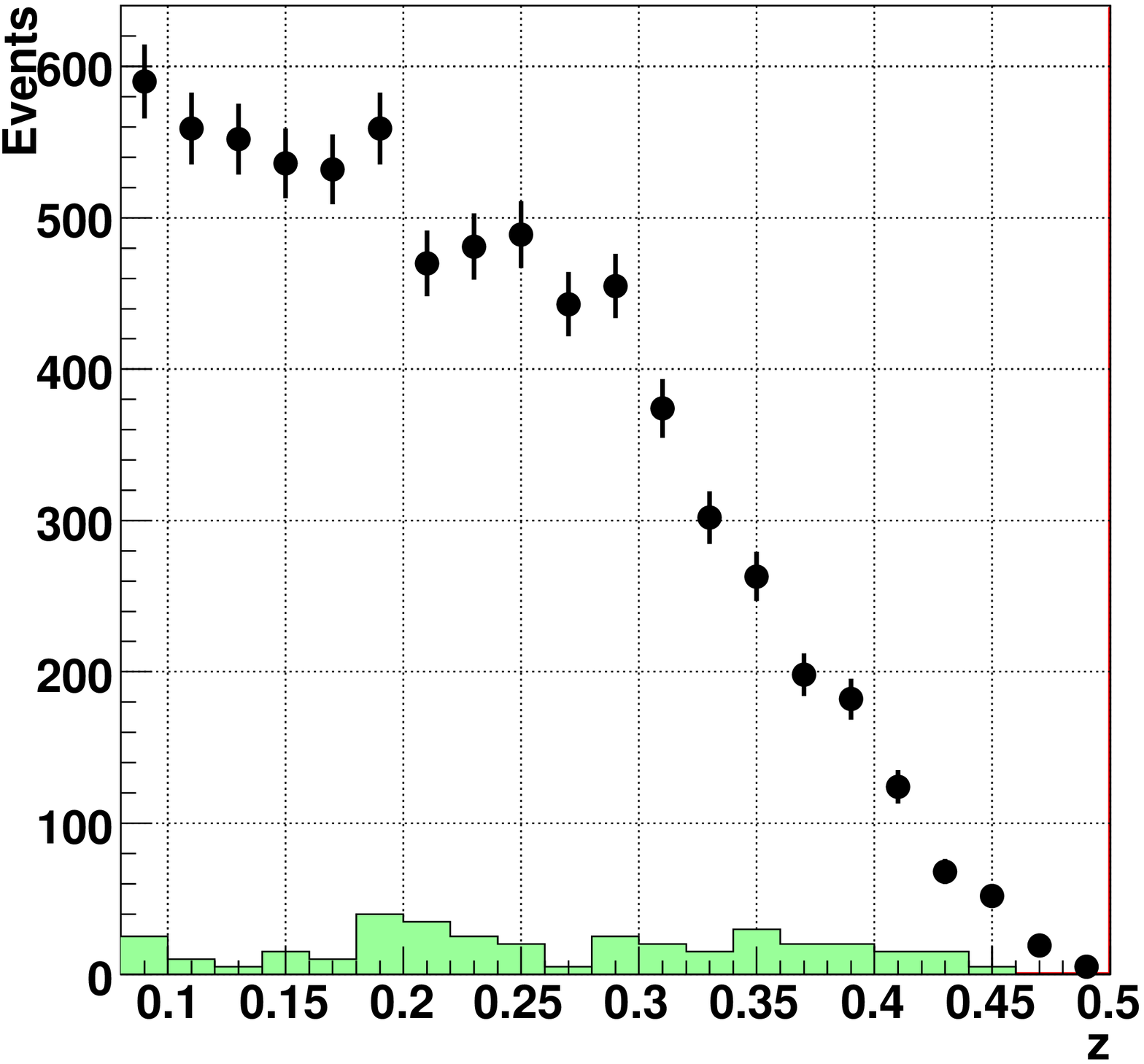}}}%
{\resizebox*{0.5\textwidth}{!}{\includegraphics{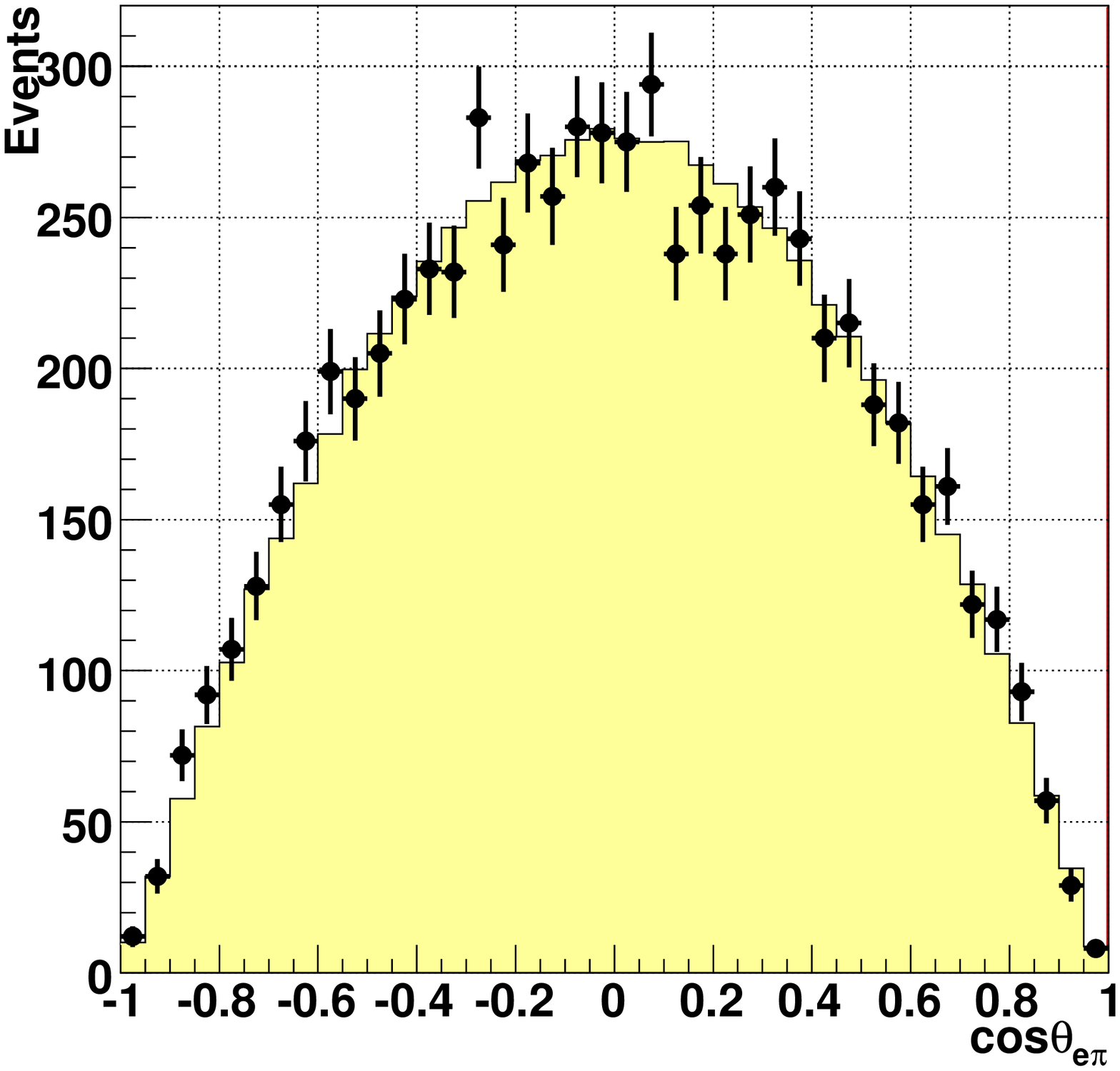}}}
\put(-257,184){\bf\large (a)} \put(-28,184){\bf\large (b)}
\end{center}
\vspace{-9mm} \caption{(a) $z$ spectrum of the selected
$K^\pm\to\pi^\pm e^+e^-$ candidates. Filled area: background
estimated with same lepton sign and $|Q|=3$ events, multiplied by a
factor of 5 for visibility. (b) $\theta_{\pi e}$ spectrum of the
selected candidates (dots) and its comparison to MC assuming vector
interaction (filled area).} \label{fig:z}
\end{figure}

\noindent The observed spectra of the data events in $z$ (in the
visible region $z>0.08$) and $\cos\theta_{\pi e}$ variables are
presented in Fig.~\ref{fig:z}. The latter spectrum, which contains
no information about $W(z)$, is compared to the MC expectation,
demonstrating a good agreement.

\begin{figure}[t]
\begin{center}
{\resizebox*{0.5\textwidth}{!}{\includegraphics{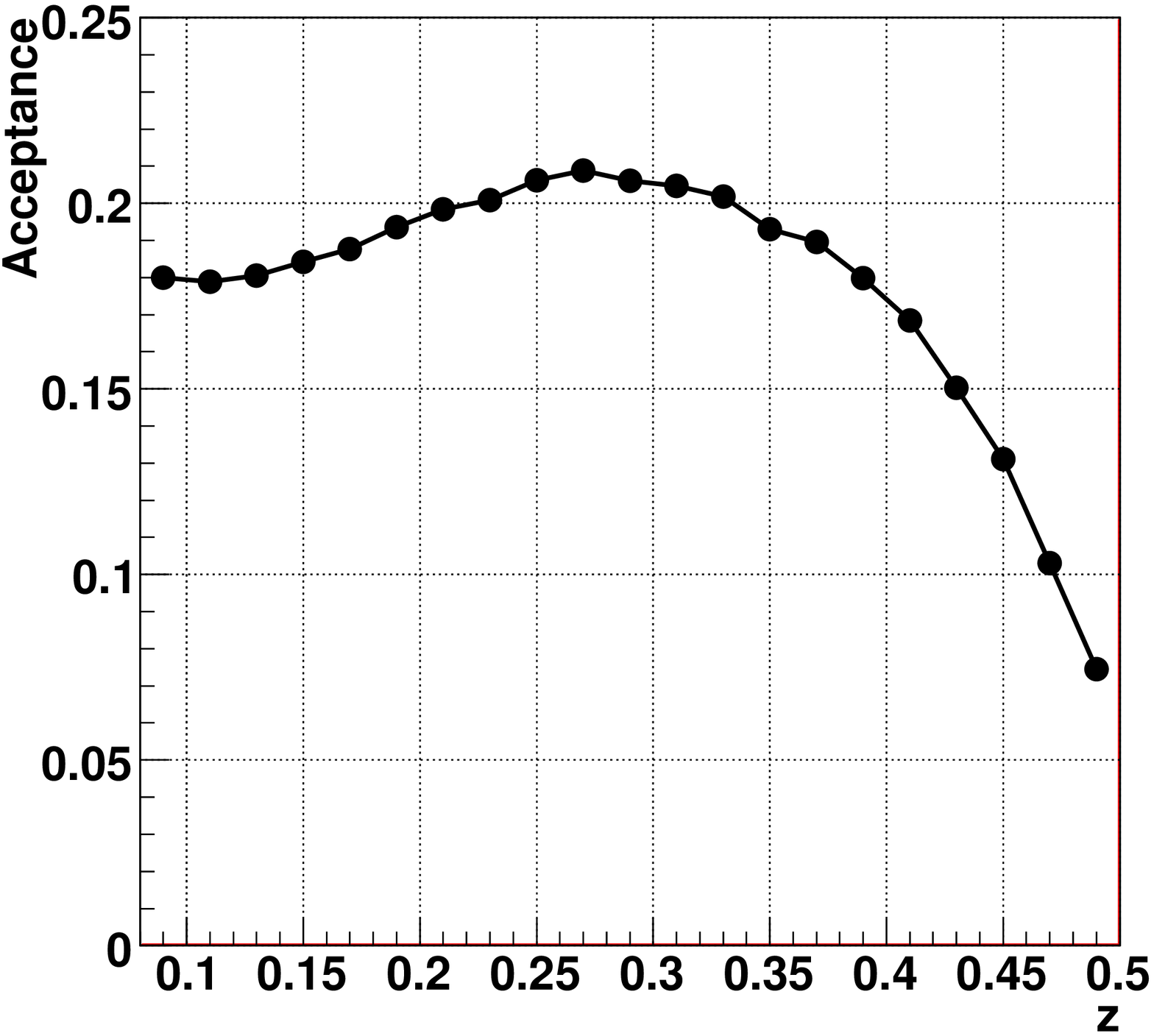}}}%
{\resizebox*{0.5\textwidth}{!}{\includegraphics{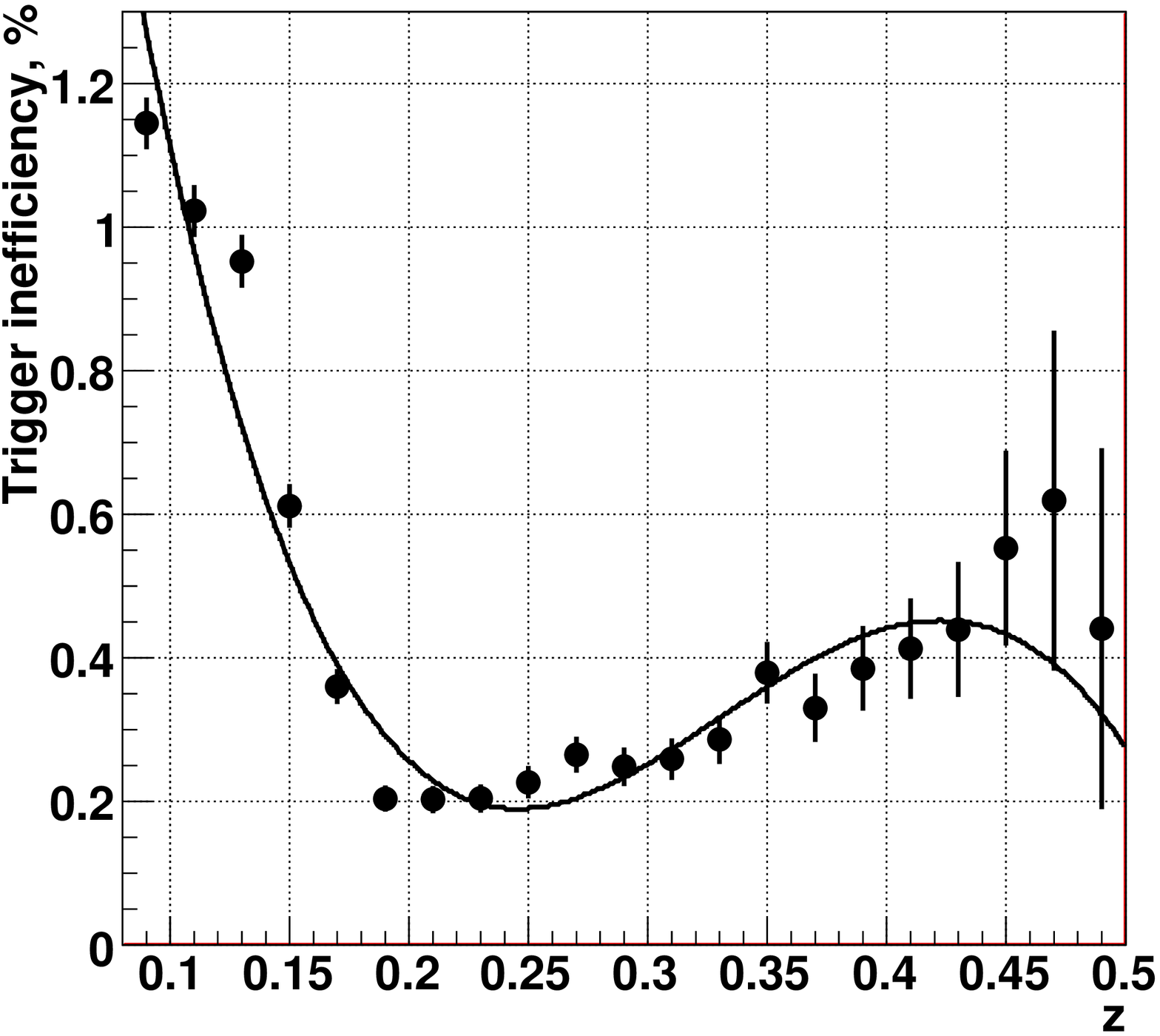}}}
\put(-257,178){\bf\large (a)} \put(-30,178){\bf\large (b)}
\end{center}
\vspace{-9mm} \caption{(a) Geometrical acceptances $A_i$ and (b)
trigger inefficiencies $\varepsilon_i$ for the $K^\pm\to\pi^\pm
e^+e^-$ sample in bins of $z$, both obtained by MC simulations. The
kinematical dependence of the trigger inefficiency is mainly due to
enhancement of L1 inefficiency for event topologies with two tracks
hitting the same HOD segment. The polynomial superimposed on Figure
(b) is drawn only to guide the eye.} \label{fig:acc}
\end{figure}

The values of $d\Gamma_{\pi ee}/dz$ in the centre of each $i$-bin of
$z$, which can be directly compared to the theoretical predictions
(\ref{theory}), are then computed as
\begin{equation}
(d\Gamma_{\pi ee}/dz)_i = \frac{N_i-N^B_i}{N_{2\pi}}\cdot
\frac{A_{2\pi}(1-\varepsilon_{2\pi})}{A_i(1-\varepsilon_i)} \cdot
\frac{1}{\Delta z} \cdot \frac{\hbar}{\tau_K} \cdot {\rm
BR}(K^\pm\to\pi^\pm\pi^0)\cdot{\rm BR}(\pi^0_D). \label{dgdz}
\end{equation}
Here $N_i$ and $N^B_i$ are the numbers of observed $K^\pm\to\pi^\pm
e^+e^-$ candidates and background events in the $i$-th bin,
$N_{2\pi}$ is the number of $K^\pm\to\pi^\pm\pi^0_D$ events (with
background subtracted), $A_i$ and $\varepsilon_i$ are geometrical
acceptance and trigger inefficiency in the $i$-th bin for the signal
sample (computed by MC simulation, both presented in
Fig.~\ref{fig:acc}), $A_{2\pi}=2.95\%$ and
$\varepsilon_{2\pi}=1.17\%$ are those for $K^\pm\to\pi^\pm\pi^0_D$
events, $\Delta z$ is the $z$ bin width, and is set to 0.02. The
external inputs are the kaon lifetime
$\tau_K=(1.2380\pm0.0021)\times10^{-8}$~s, and branching ratios of
the normalisation decay modes ${\rm
BR}(K^\pm\to\pi^\pm\pi^0)=(20.68\pm0.13)\%$, ${\rm
BR}(\pi^0_D)=(1.198\pm0.032)\%$~\cite{pdg}.

While the simulated values of geometric acceptances and trigger
efficiencies are subject to relatively large systematic
uncertainties (discussed in the following section), their ratios
$A_i/A_{2\pi}$ and $(1-\varepsilon_i)/(1-\varepsilon_{2\pi})$
appearing in (\ref{dgdz}) are affected by these uncertainties to a
lesser extent due to cancellations of most systematic effects.

%The kaon decay flux in the fiducial volume is computed to be
%\begin{equation}
%\Phi_K=N_{2\pi}/{\rm BR}(K^\pm\to\pi^\pm\pi^0)/{\rm
%BR}(\pi^0_D)/A_{2\pi}/(1-\varepsilon_{2\pi})=(1.698\pm0.046)\times
%10^{11},
%\end{equation}
%where the only the $2.7\%$ relative uncertainty due to external
%input is indicated.

\begin{figure}[t]
\begin{center}
{\resizebox*{0.5\textwidth}{!}{\includegraphics{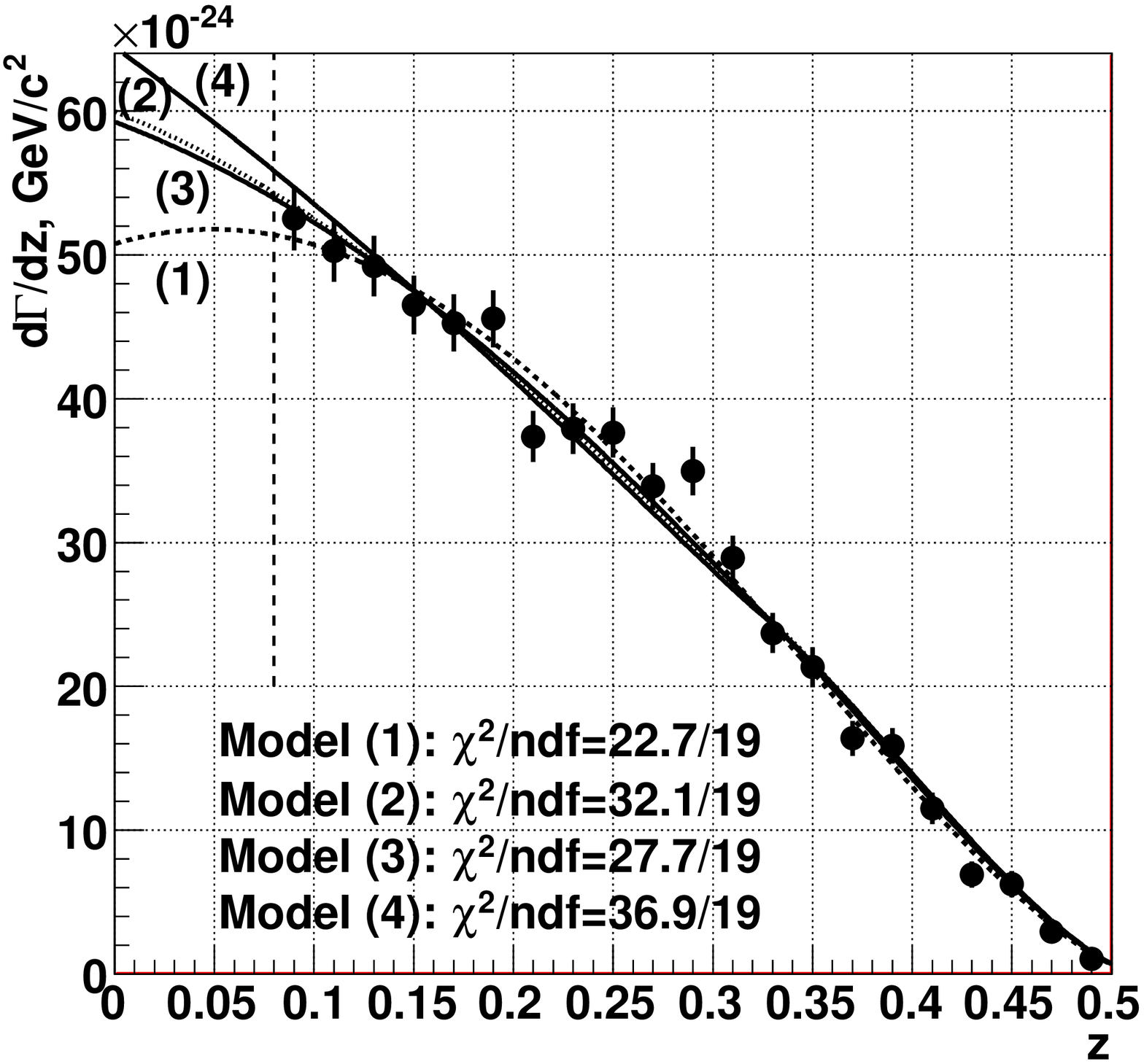}}}%
{\resizebox*{0.5\textwidth}{!}{\includegraphics{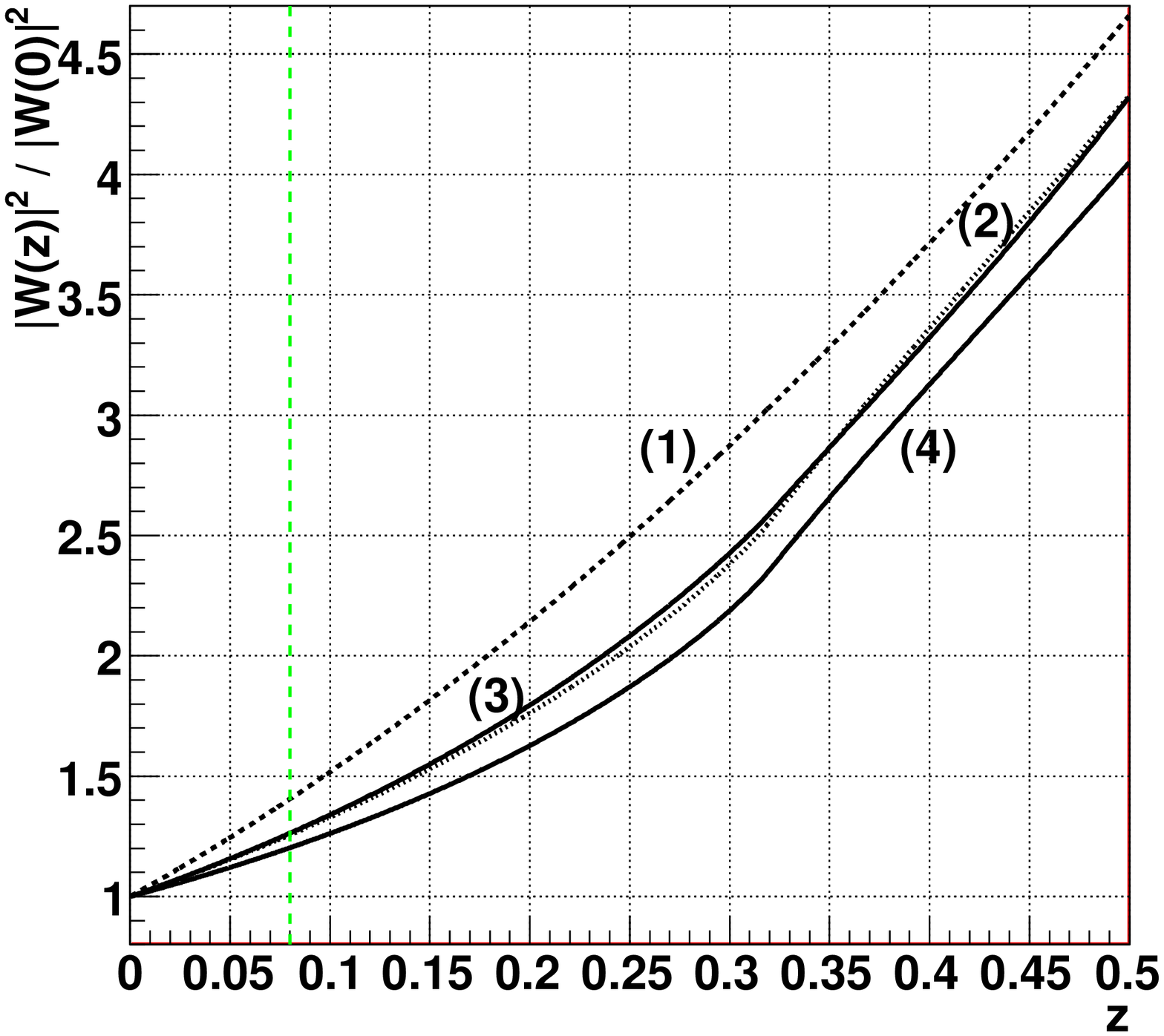}}}
\put(-260,188){\bf\large (a)} \put(-45,188){\bf\large (b)}
\end{center}
\vspace{-9mm} \caption{(a) $d\Gamma_{\pi ee}/dz$ (background
subtracted, corrected for trigger efficiency) and fit results
according to the four considered models. (b) Squared form factors
$|W(z)|^2$ for each model normalized by a condition $|W(0)|=1$.}
\label{fig:fit}
\end{figure}

The computed values of $d\Gamma_{\pi ee}/dz$ vs $z$ and the results
of the fits to the four considered models are presented in
Fig.~\ref{fig:fit}a. The corresponding squared form factors
$|W(z)|^2$ normalized by a condition $|W(0)|=1$ are presented in
Fig.~\ref{fig:fit}b.

Branching ratios ${\rm BR}(K^\pm\to\pi^\pm e^+e^-)$ in the full
kinematic range corresponding to each model are computed using the
measured value of the model parameters, their statistical
uncertainties, and correlation matrices. In addition, a
model-independent branching ratio ${\rm BR_{mi}}$ in the visible
kinematic region $z>0.08$ is computed by integration of
$d\Gamma_{\pi ee}/dz$. ${\rm BR_{mi}}$ differs from each of the
model-dependent BRs computed in the visible $z$ range by less than
$0.01\times10^{-7}$. The differences between the model-dependent BRs
come from the region $z<0.08$, as can be seen in
Fig.~\ref{fig:fit}a.

\vspace{2mm}

\noindent {\bf Systematic uncertainties} \vspace{2mm}

\noindent {\bf Particle identification.} Imperfect MC description of
electron and pion identification inefficiencies $f_e$ and $f_\pi$
can bias the result only due to the momentum dependence of $f_e$ and
$f_\pi$, due to identical charged particle composition, but
differing $\pi$ and $e$ momentum spectra of the signal and
normalisation modes. Inefficiencies were measured with the data to
be in the ranges $1.6\%<f_\pi<1.7\%$ and $1.1\%<f_e<1.7\%$,
depending in momentum. Systematic uncertainties due to these
momentum dependencies not perfectly described by MC were
conservatively estimated assuming momentum-independent MC $f_e$ and
$f_\pi$.

{\bf Beam line description.} Despite the careful simulation of the
beamline including time variations of its parameters, the residual
discrepancies of data and MC beam geometries and spectra may bias
the results. To evaluate the related systematic uncertainties,
variations of the results with respect to variations of cuts on
track momenta, LKr cluster energies, total and transverse momenta of
the $\pi^\pm e^+e^-(\gamma)$ final states, track distances from beam
axis in DCH planes, and distances between particle impact points at
the LKr calorimeter surface were studied. Stability of the results
in bins of decay vertex longitudinal coordinate was studied as well.
The maximum observed variations of the fit parameters are taken
conservatively as the systematic uncertainties.

{\bf Background subtraction.} For background subtraction in the
$\pi^\pm e^+e^-$ sample the spectra of same lepton sign and $|Q|=3$
events were used, as described above. This method has a limited
statistical precision (with an average of $\sim 3$ background events
in a bin of $z$). The uncertainties of the measured parameters due
to background subtraction were conservatively taken to be equal to
the corrections themselves.

{\bf Trigger efficiency.} As discussed earlier, the corrections for
trigger inefficiencies were evaluated by simulations. In particular,
L1 and L2 corrections for the measured BR have similar magnitudes of
a few $10^{-3}$. No uncertainty was assigned to the L1 correction,
due to relative simplicity of the trigger condition, and the
consequent robustness of the simulation. On the other hand, the
uncertainty of the L2 efficiency correction was conservatively taken
to be equal to the correction itself.

{\bf Radiative corrections.} Uncertainties due to the radiative
corrections were evaluated by variation of the lower $\pi^\pm
e^+e^-$ invariant mass cut. The results show reasonable stability
with respect to the variation, as can be expected given a good MC
description of the $\pi^\pm e^+e^-$ radiative mass tail visible in
Fig.~\ref{fig:mk}a. Estimation of systematic uncertainties as
differences of fit results obtained with radiative corrections
calculated using the PHOTOS simulation and the formalism given
in~\cite{is08} leads to similar results as those obtained with the
first method.

{\bf Fitting method.} Uncertainties due to the fitting procedure
were evaluated by comparing the fit results obtained with a bin
width $\Delta z=0.01$ with those obtained with the standard bin
width of $\Delta z=0.02$.

{\bf External input.} Substantial uncertainties arise from the
external input, as ${\rm BR}(\pi^0_D)=(1.198\pm0.032)\%$ is
experimentally known with a limited relative precision of
2.7\%~\cite{pdg}. The only parameter not affected by this external
uncertainty is the slope $\delta$ of the linear form factor
describing the shape of the spectrum.

\begin{table}[tb]
\begin{center}
\caption{Corrections and systematic uncertainties (excluding
external sources).} \vspace{-2mm}
\begin{tabular}{@{}r@{~~}r@{~~}r@{~~}r@{~$\pm$~}lr@{~$\pm$~}l@{~~}r@{~~}r}
\hline Parameter&Particle&Beam      &\multicolumn{2}{c}{Background} &\multicolumn{2}{c}{Trigger}&Radiative&Fitting\\
                &ID      &simulation&\multicolumn{2}{c}{subtraction}&\multicolumn{2}{c}{inefficiency}   &corr.    &method\\
\hline
$|f_0|$        &0.001&0.006&$0.001$ &0.001&$0.002$ &0.002&0.004&0.003\\
$\delta$       & 0.01& 0.04&$-0.05$ &0.05 &$-0.03$ &0.03 & 0.04&0.03\\
$a_+$          &0.001&0.005&$-0.001$&0.001&$-0.002$&0.002&0.004&0.004\\
$b_+$          &0.009&0.015&$0.021$ &0.021&$0.015$ &0.015&0.014&0.010\\
$\tilde{\rm w}$&0.001&0.002&$-0.002$&0.002&$-0.001$&0.001&0.002&0.001\\
$\beta$        & 0.06& 0.09&$-0.09$ &0.09 &$-0.06$ &0.06 &0.06&0.04\\
$M_a$/GeV&0.004&0.009&$0.009$ &0.009&$0.007$ &0.007&0.009&0.006\\
$M_b$/GeV&0.002&0.003&$0.004$ &0.004&$0.003$ &0.003&0.004&0.002\\
\hline
${\rm BR}_{1-4}\times 10^7$&0.02&0.02&$-0.02$&0.02&$-0.01$&0.01&0.02&0.02\\
${\rm BR_{mi}}\times 10^7$     &0.02&0.02&$-0.01$&0.01&$-0.01$&0.01&0.02&n/a\\
\hline
\end{tabular}
\end{center}
\vspace{-4mm} \label{tab:syst}
\end{table}

The applied corrections and the systematic uncertainties, excluding
those from external sources, are summarized in Table 1.
Uncertainties due to the external input can be found in Table 2.

\section{Results and discussion}

\begin{figure}[tb]
\begin{center}
\vspace{+1mm}
\begin{tabular}{cc}
{\resizebox*{0.4\textwidth}{!}{\includegraphics{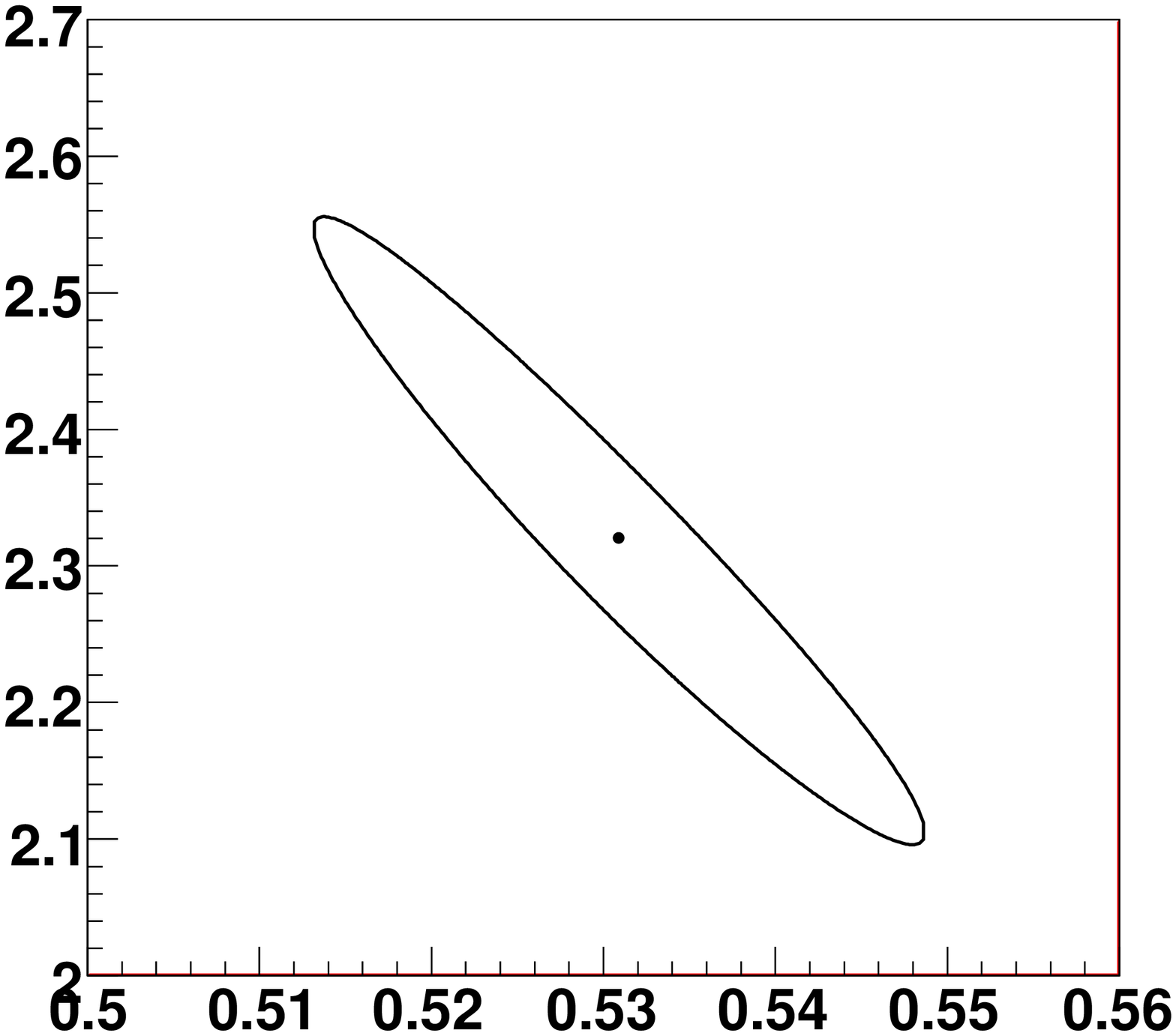}}}
\put(-160,148){\bf\large $\delta$} \put(-32,25){\large $|f_0|$}&
{\resizebox*{0.4\textwidth}{!}{\includegraphics{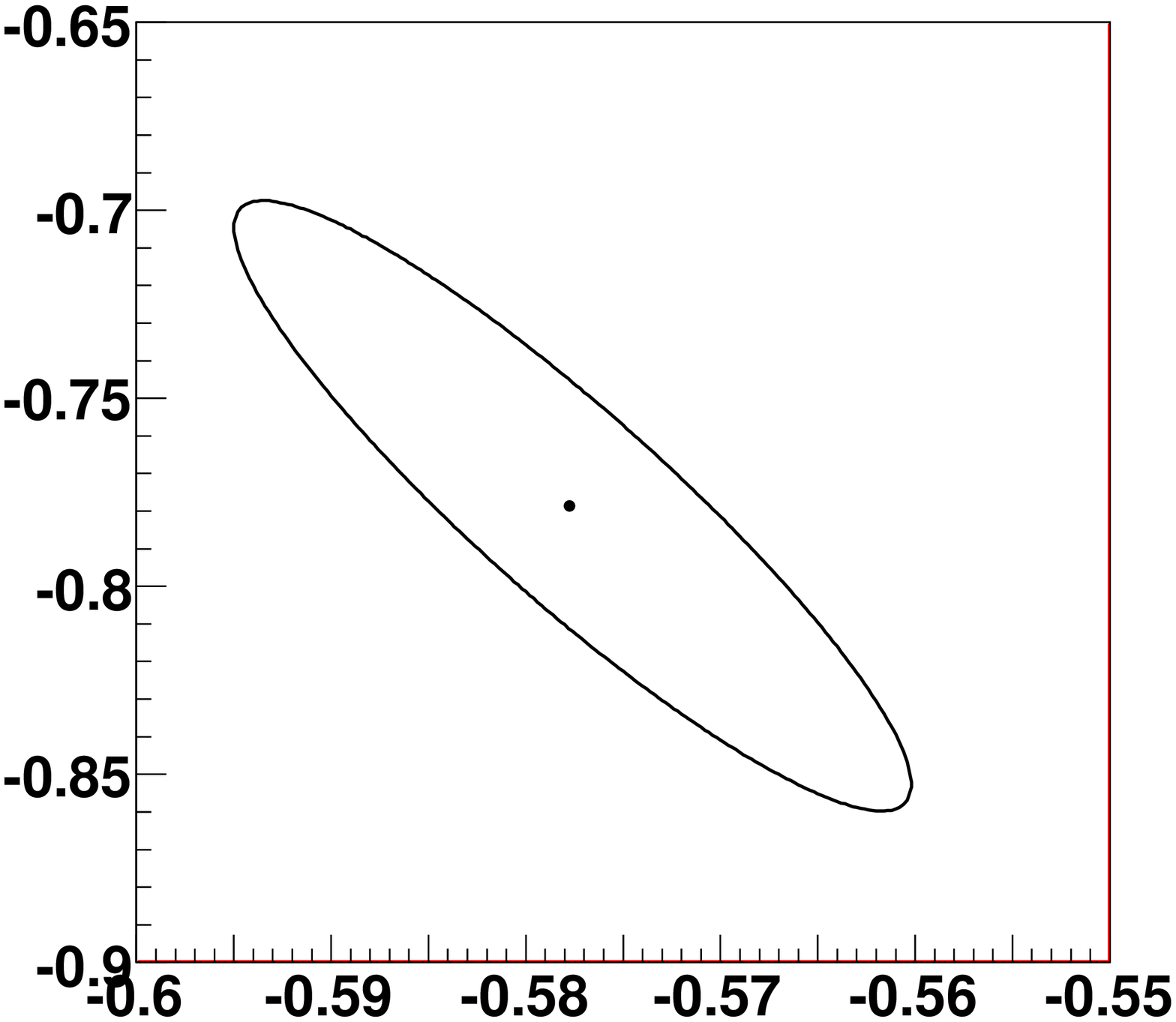}}}
\put(-156,148){\bf\large $b_+$} \put(-30,25){\large $a_+$}\\
{\resizebox*{0.4\textwidth}{!}{\includegraphics{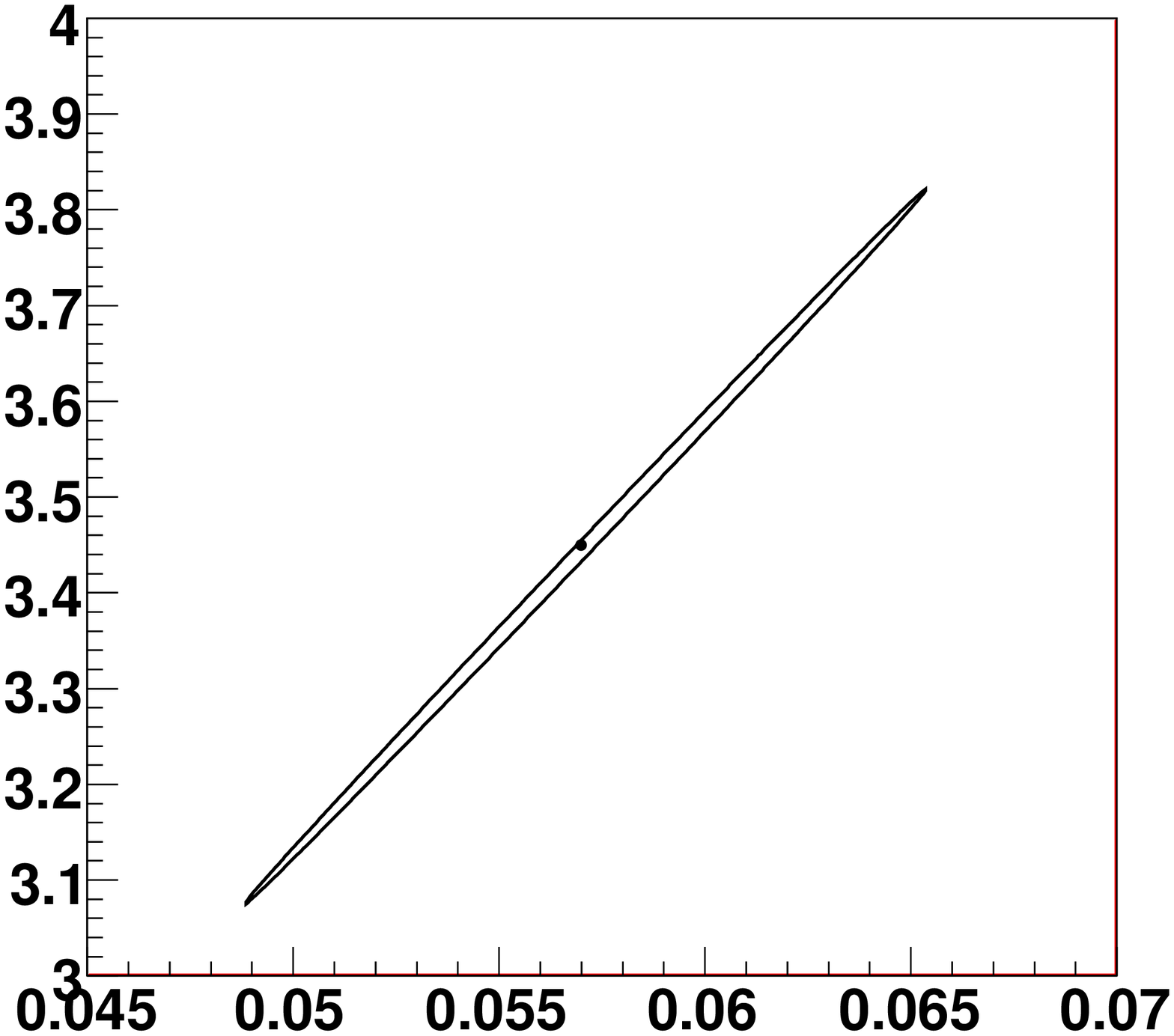}}}
\put(-160,148){\bf\large $\beta$} \put(-25,25){\large $\tilde{\rm
w}$}&
{\resizebox*{0.4\textwidth}{!}{\includegraphics{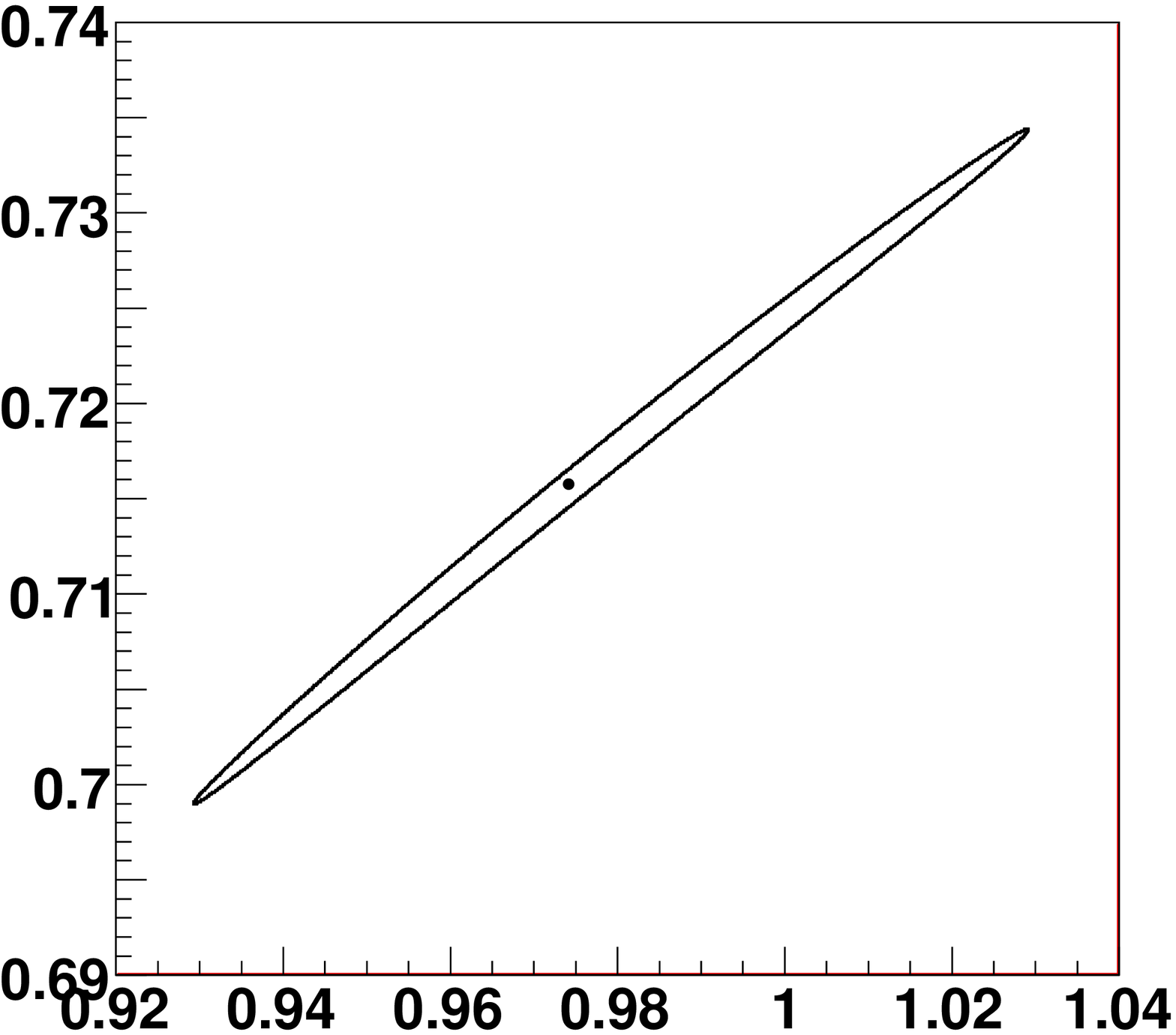}}}
\put(-158,148){\large $M_\rho/({\rm GeV}/c^2)$}
\put(-95,25){\large $M_a/({\rm GeV}/c^2)$}\\
\end{tabular}
\end{center}
\vspace{-9mm} \caption{68\% confidence level contours for the
parameters of the four models. Only statistical and background
subtraction errors are included.} \label{fig:cl}
\end{figure}

\noindent The measured value of the model-independent ${\rm
BR_{mi}}(z>0.08)$, as well as the parameters of the considered
models and the corresponding BRs in the full $z$ range, with their
statistical, systematic, and external uncertainties are presented in
Table 2. The 68\% confidence level contours for the pairs of
parameters corresponding to each model are presented in
Fig.~\ref{fig:cl}. The corresponding correlation coefficients
between the model parameters are $\rho(|f_0|,\delta)=-0.962$,
$\rho(a_+,b_+)=-0.913$, $\rho(\tilde{\rm w},\beta)=0.999$ and
$\rho(M_a,M_\rho)=0.998$.

\begin{table}[tb]
 \label{tab:results}
\begin{center}
\caption{Model-independent ${\rm BR_{mi}}(z>0.08)$, and fit results
for the considered models.}
\begin{tabular}{rrrrrrrrrrrr}
\hline
${\rm BR_{mi}}\times10^7=\!\!\!$       &$\!2.28$  &$\!\!\pm\!\!$&$0.03_{\rm stat.}$ &$\!\!\pm\!\!$&$0.04_{\rm syst.}$ &$\!\!\pm\!\!$&$0.06_{\rm ext.}$ &$\!\!=\!\!$&$2.28$ &$\!\!\pm\!\!$&0.08\\
\hline Model (1)$\!\!\!$\\
$|f_0|=\!\!\!$                         &$\!0.531$ &$\!\!\pm\!\!$&$0.012_{\rm stat.}$&$\!\!\pm\!\!$&$0.008_{\rm syst.}$&$\!\!\pm\!\!$&$0.007_{\rm ext.}$&$\!\!=\!\!$&$0.531$&$\!\!\pm\!\!$&0.016\\
$\delta=\!\!\!$                        &$\!2.32$  &$\!\!\pm\!\!$&$0.15_{\rm stat.}$ &$\!\!\pm\!\!$&$0.09_{\rm syst.}$ &             &                  &$\!\!=\!\!$&$2.32$ &$\!\!\pm\!\!$&0.18\\
${\rm BR}_1\times10^7=\!\!\!$          &$\!3.05$  &$\!\!\pm\!\!$&$0.04_{\rm stat.}$ &$\!\!\pm\!\!$&$0.05_{\rm syst.}$ &$\!\!\pm\!\!$&$0.08_{\rm ext.}$ &$\!\!=\!\!$&$3.05$ &$\!\!\pm\!\!$&0.10\\
\hline Model (2)$\!\!\!$\\
$a_+=\!\!\!$                           &$\!-0.578$&$\!\!\pm\!\!$&$0.012_{\rm stat.}$&$\!\!\pm\!\!$&$0.008_{\rm syst.}$&$\!\!\pm\!\!$&$0.007_{\rm ext.}$&$\!\!=\!\!$&$-0.578$&$\!\!\pm\!\!$&0.016\\
$b_+=\!\!\!$                           &$\!-0.779$&$\!\!\pm\!\!$&$0.053_{\rm stat.}$&$\!\!\pm\!\!$&$0.036_{\rm syst.}$&$\!\!\pm\!\!$&$0.017_{\rm ext.}$&$\!\!=\!\!$&$-0.779$&$\!\!\pm\!\!$&0.066\\
${\rm BR}_2\times10^7=\!\!\!$          &$\!3.14$  &$\!\!\pm\!\!$&$0.04_{\rm stat.}$ &$\!\!\pm\!\!$&$0.05_{\rm syst.}$ &$\!\!\pm\!\!$&$0.08_{\rm ext.}$ &$\!\!=\!\!$&$3.14$ &$\!\!\pm\!\!$&0.10\\
\hline Model (3)$\!\!\!$\\
$\tilde{\rm w}=\!\!\!$                 &$\!0.057$ &$\!\!\pm\!\!$&$0.005_{\rm stat.}$&$\!\!\pm\!\!$&$0.004_{\rm syst.}$&$\!\!\pm\!\!$&$0.001_{\rm ext.}$&$\!\!=\!\!$&$0.057$&$\!\!\pm\!\!$&0.007\\
$\beta=\!\!\!$                         &$\!3.45$  &$\!\!\pm\!\!$&$0.24_{\rm stat.}$ &$\!\!\pm\!\!$&$0.17_{\rm syst.}$ &$\!\!\pm\!\!$&$0.05_{\rm ext.}$ &$\!\!=\!\!$&$3.45$ &$\!\!\pm\!\!$&0.30\\
${\rm BR}_3\times10^7=\!\!\!$          &$\!3.13$  &$\!\!\pm\!\!$&$0.04_{\rm stat.}$ &$\!\!\pm\!\!$&$0.05_{\rm syst.}$ &$\!\!\pm\!\!$&$0.08_{\rm ext.}$ &$\!\!=\!\!$&$3.13$ &$\!\!\pm\!\!$&0.10\\
\hline Model (4)$\!\!\!$\\
$M_a/{\rm GeV}/c^2)=\!\!\!$            &$\!0.974$ &$\!\!\pm\!\!$&$0.030_{\rm stat.}$&$\!\!\pm\!\!$&$0.019_{\rm syst.}$&$\!\!\pm\!\!$&$0.002_{\rm ext.}$&$\!\!=\!\!$&$0.974$&$\!\!\pm\!\!$&0.035\\
$M_\rho/({\rm GeV}/c^2)=\!\!\!$        &$\!0.716$ &$\!\!\pm\!\!$&$0.011_{\rm stat.}$&$\!\!\pm\!\!$&$0.007_{\rm syst.}$&$\!\!\pm\!\!$&$0.002_{\rm ext.}$&$\!\!=\!\!$&$0.716$&$\!\!\pm\!\!$&0.014\\
${\rm BR}_4\times10^7=\!\!\!$          &$\!3.18$  &$\!\!\pm\!\!$&$0.04_{\rm stat.}$ &$\!\!\pm\!\!$&$0.05_{\rm syst.}$ &$\!\!\pm\!\!$&$0.08_{\rm ext.}$ &$\!\!=\!\!$&$3.18$ &$\!\!\pm\!\!$&0.10\\
\hline
\end{tabular}
\end{center}
\vspace{-4mm}
\end{table}

The measurement of the model-independent ${\rm BR_{mi}}(z>0.08)$,
while not involving extrapolation into the region $z<0.08$ depending
on the form factor parameterization, still relies on the
extrapolation into the region of the radiative tail $M_{\pi
ee}<470$~MeV/$c^2$.

Each of the four considered models provides a reasonable fit to the
data ($\chi^2$ of the fits are indicated in Fig.~\ref{fig:fit}a),
however the linear form factor model leads to the smallest $\chi^2$.
The size of the data sample is insufficient to distinguish between
the models.

Earlier measurements of the decay parameters are summarized in Table
3. The present measurement of the linear form factor slope $\delta$
is in agreement with the previous ones based on
$K^+\to\pi^+e^+e^-$~\cite{al92,ap99} and
$K^+\to\pi^+\mu^+\mu^-$~\cite{ma00} samples, and further confirms
the contradiction of the data to meson dominance models~\cite{li99}
which predict lower values of the slope parameter. The obtained
$|f_0|$, $a_+$ and $b_+$ are in agreement with the only previous
measurement~\cite{ap99}. The obtained values of $\tilde{\rm w}$ and
$\beta$ are in fair agreement with the fit of the data
points~\cite{ap99} performed by the authors of~\cite{fr04} (the
value of $M_\rho=0.77549$~GeV/$c^2$~\cite{pdg} is used in our fit;
using $M_\rho=0.7711$~GeV/$c^2$ as in~\cite{fr04} leads to a central
value of $\beta=3.27$, which improves agreement with~\cite{fr04}).
The measured parameter $M_a$ is consistent with the nominal mass of
the $a_0(980)$ resonance, while $M_\rho$ is a few \% lower than the
nominal $\rho(770)$ mass~\cite{pdg}.

\begin{table}[tb]
\label{tab:previous}
\begin{center}
\caption{Summary of earlier measurements of the form factor
parameters and BR with their statistical, systematic, and
model-dependence uncertainties. Systematic uncertainties were not
evaluated for some measurements; a single statistical uncertainty is
presented for these cases.}\vspace{1mm}
\begin{tabular}{ccccc}
\hline
Decay & $K_{\pi ee}^+$ & $K_{\pi ee}^+$ & $K_{\pi ee}^+$ & $K_{\pi\mu\mu}^+$\\
Refs. & \cite{bl75} & \cite{al92} & \cite{ap99,fr04} & \cite{ma00}\\
\hline
$\delta$       &&$1.31\pm0.44\pm0.19$&$2.14\pm0.13\pm0.15$&$2.45_{-0.95}^{+1.30}$\\
$|f_0|$        &&                &$0.533\pm0.012$\\
$a_+$          &&                &$-0.587\pm0.010$\\
$b_+$          &&                &$-0.655\pm0.044$\\
$\tilde{\rm w}$&&                &$0.045\pm0.003$\\
$\beta$        &&                &$2.8\pm0.1$\\
${\rm BR}\times 10^7$ & $2.7\pm0.5$ &$2.75\pm0.23\pm0.13$ & $2.94\pm0.05\pm0.13\pm0.05$\\
\hline
\end{tabular}
\end{center}
\vspace{-4mm}
\end{table}

The branching ratio in the full kinematic range, which is computed
as the average between the two extremes corresponding to the models
(1) and (4), and includes an uncertainty due to extrapolation into
the inaccessible region $z<0.08$, is
\begin{displaymath}
{\rm BR}=(3.11\pm0.04_{\rm stat.}\pm0.05_{\rm syst.}\pm0.08_{\rm
ext.}\pm0.07_{\rm model})\times10^{-7}= (3.11\pm0.12)\times10^{-7}.
\end{displaymath}
The uncertainty of the above result is correlated with those of the
earlier measurements (those due to external input and
model-dependence uncertainties). A comparison to the most precise of
these measurements by the BNL E865 collaboration~\cite{ap99} using
the same external input, and taking into account correlation of
external uncertainties between the two analyses, shows a $1.6\sigma$
level of agreement. In conclusion, the obtained BR is in agreement
with the previous measurements~\cite{bl75, al92, ap99}.

Measurements of the decay parameters and BRs were performed
separately for $K^+$ and $K^-$ decays. In particular, the branching
fractions in the full kinematic range and their statistical
uncertainties were measured to be
\begin{displaymath}
{\rm BR}^+=(2.99\pm0.05_{\rm stat.})\times10^{-7},~~~{\rm
BR}^-=(3.13\pm0.08_{\rm stat.})\times10^{-7}.
\end{displaymath}
This allows making a first measurement of the direct CP violating
asymmetry of $K^+$ and $K^-$ decay rates in the full kinematic
range. Considering only the uncorrelated systematic uncertainties
(those due to background subtraction) between $K^+$ and $K^-$
samples,
\begin{displaymath}
\Delta(K_{\pi ee}^\pm)=({\rm BR}^+-{\rm BR}^-)/({\rm BR}^++{\rm
BR}^-)=(-2.2\pm1.5_{\rm stat.}\pm 0.6_{\rm syst.})\times 10^{-2}.
\end{displaymath}
The other systematic uncertainties are correlated between the $K^+$
and $K^-$ samples, and thus cancel in $\Delta(K_{\pi ee}^\pm)$. A
conservative limit for the charge asymmetry of $|\Delta(K_{\pi
ee}^\pm)|<2.1\times10^{-2}$ at 90\% CL can be deduced from the above
value. This result is compatible with CP conservation, however the
achieved precision is far from the Standard Model expectation
$|\Delta(K_{\pi ee}^\pm)|\sim 10^{-5}$~\cite{da98} and even the SUSY
upper limit $|\Delta(K_{\pi ee}^\pm)|\sim 10^{-3}$~\cite{me02,da02}
for the CP violating asymmetry.

\section*{Summary}

From a sample of 7253 $K^\pm\to\pi^\pm e^+e^-$ decay candidates with
1.0\% background contamination, the branching fraction in the full
kinematic range has been measured to be
$(3.11\pm0.12)\times10^{-7}$, in agreement and competitive with
previous measurements. The shape of the form factor which
characterizes the decay has been evaluated in the framework of four
models, giving consistent results with previous measurements. The
first simultaneous observation of both charge kaon decays into
$\pi^\pm e^+e^-$ allowed to establish an upper limit for the CP
violating asymmetry of $K^+$ and $K^-$ decay rates of $2.1\times
10^{-2}$ at 90\% CL.

%%%%%%%%%%%%%%%%%%%%%%%%%%%%%%%%%
\section*{Acknowledgements}

We gratefully acknowledge the staff of the CERN SPS accelerator and
the beamline for the excellent performance of the beam. We also
thank the technical staff of the participating laboratories,
universities and affiliated computing centres for their efforts in
operation of the experiment and data processing. We are grateful to
Samuel Friot, Gino Isidori and Victor Pervushin for valuable
discussions.

%%%%%%%%%%%%%%%%%%%%%%%%%%%%%%%%%
\section*{Appendix}

The paper~\cite{fr04} that formulates one of the considered models
does not contain the full definition for the form factor $f_V(z)$:
the function $\chi(z)$ appearing in the Eq. (3.19) for $f_V(z)$ is
defined only below the $2\pi$ threshold. The expressions for
$\chi(z)$ both below and above the threshold, kindly provided by the
authors of Ref.~\cite{fr04}, are presented below. Let us define
$r=(M_K/m_\pi)^2$. Then for $z<4/r$ (below the threshold),
\begin{displaymath}
\chi(z) =
\frac{4}{9}-\frac{4}{3zr}+\frac{1}{3}\left(\frac{4}{zr}-1\right)^{3/2}
\arctan\left(\frac{1}{\sqrt{\frac{4}{zr}-1}}\right),
\end{displaymath}
and for $z\ge4/r$ (above the threshold),
\begin{displaymath}
\chi(z) = \frac{4}{9}-\frac{4}{3zr}+\frac{1}{6}
\left(1-\frac{4}{zr}\right)^{3/2} \left\{ \ln \left(
\frac{1-\sqrt{1-\frac{4}{zr}}}{1+\sqrt{1-\frac{4}{zr}}} \right)
-i\pi \right\}.
\end{displaymath}

%%%\end{linenumbers}
%%%%%%%%%%%%%%%%%%%%%%%%%%%%%%%%%%%%%%%%%%%%

\end{document}